\documentclass[a4paper,10pt]{article}
\usepackage[english]{babel}
\usepackage[utf8]{inputenc}
\usepackage[nottoc]{tocbibind}
\usepackage[affil-it]{authblk}
\usepackage{amsmath}%
\usepackage{amsfonts}%
\usepackage{amssymb}%
\usepackage[a4paper]{geometry}
\usepackage{graphicx}
\usepackage{mathrsfs}
\usepackage{bm}
\usepackage{bbold}
\usepackage{slashed}
\usepackage[hidelinks]{hyperref}
\usepackage{mathtools}
\usepackage{csquotes}
\usepackage{biblatex}
\hypersetup{
   colorlinks = true,
    linkcolor = {black},
    anchorcolor = {green},
    citecolor = {red},
    filecolor = {cyan},
    menucolor = {red},
    runcolor = {cyan},
    urlcolor = {magenta}
}

\newcommand{\II}{{\mathcal{I}}}

\newcommand{\CC}{{\mathcal{C}}}
\newcommand{\GG}{{\mathcal{G}}}

\newcommand{\MM}{{\mathcal{M}}}

\newcommand{\R}{{\mathbb{R}}}

\newcommand{\Z}{{\mathbb{Z}}}
\newcommand{\N}{{\mathbb{N}}}
\newcommand{\C}{{\mathbb{C}}}

\newcommand{\CP}{{\mathbb{C}}{{\mathbb{P}}}}

\newcommand{\beq}{\begin{equation}}
\newcommand{\eeq}{\end{equation}}
\newcommand{\bea}{\begin{eqnarray}}
\newcommand{\eea}{\end{eqnarray}}
\newcommand{\bal}{\begin{align}}
\newcommand{\eal}{\end{align}}

\newcommand{\Img}{{\mathrm{Im}}}
\newcommand{\adj}{{\mathrm{adj}}}

\newcommand{\bdy}{\partial}

\newcommand{\wt}{\widetilde}
\newcommand{\wh}{\widehat}

\renewcommand{\to}{\longrightarrow}

\newcommand{\ol}{\overline}
\newcommand{\tr}{{\rm tr}\, }

\newcommand{\sn}{{\rm sn}\, }
\newcommand{\cn}{{\rm cn}\, }
\newcommand{\dn}{{\rm dn}\, }
\newcommand{\scn}{{\rm sc}\, }
\newcommand{\nc}{{\rm nc}\, }

\newcommand{\dc}{{\rm dc}\, }

\newcommand{\diag}{{\rm diag}}

\newcommand{\bigslant}[2]{{\raisebox{.2em}{$#1$}\left/\raisebox{-.2em}{$#2$}\right.}}

\renewcommand{\epsilon}{\varepsilon}

\newtheorem{theorem}{Theorem}

\newtheorem{corollary}[theorem]{Corollary}

\newtheorem{definition}[theorem]{Definition}

\newtheorem{lemma}[theorem]{Lemma}
\newtheorem{notation}[theorem]{Notation}

\newtheorem{proposition}[theorem]{Proposition}
\newtheorem{remark}[theorem]{Remark}

\makeatletter
\newcommand{\Spvek}[2][r]{%
  \gdef\@VORNE{1}
  \left(\hskip-\arraycolsep%
    \begin{array}{#1}\vekSp@lten{#2}\end{array}%
  \hskip-\arraycolsep\right)}

\def\vekSp@lten#1{\xvekSp@lten#1;vekL@stLine;}
\def\vekL@stLine{vekL@stLine}
\def\xvekSp@lten#1;{\def\temp{#1}%
  \ifx\temp\vekL@stLine
  \else
    \ifnum\@VORNE=1\gdef\@VORNE{0}
    \else\@arraycr\fi%
    #1%
    \expandafter\xvekSp@lten
  \fi}
\makeatother

\title{Symmetric calorons and the rotation map}
\author{Josh Cork}
\affil{School of Mathematics,\\
University of Leeds,\\
LS2 9JT\\
Email address: mmjsc@leeds.ac.uk}
\date{\today}
\begin{document}

\maketitle

    \begin{center}
\textbf{Abstract}
			\end{center}
\vskip0.25cm
	We study $SU(2)$ calorons, also known as periodic instantons, and consider invariance under isometries of $S^1\times\R^3$ coupled with a non-spatial isometry called the rotation map. In particular, we investigate the fixed points under various cyclic symmetry groups. Our approach utilises a construction akin to the ADHM construction of instantons -- what we call the \textit{monad matrix data} for calorons -- derived from the work of Charbonneau and Hurtubise. To conclude, we present an example of how investigating these symmetry groups can help to construct new calorons by deriving Nahm data in the case of charge $2$.
\vskip0.5cm
\hypersetup{
    linkcolor = {blue}
}
\section{Introduction}
For quite some time, there has been great interest in the various variational problems which arise in physical field theories, in particular those that yield topological soliton solutions \cite{MantonSutcliffe2004}. Examples of these are the Yang-Mills, Yang-Mills-Higgs, and Skyrme theories, whose soliton solutions are called instantons, monopoles, and skyrmions respectively. The topological aspect of these objects is usually due to them being classified by an  \textit{integral charge}, which does not change under continuous deformations, and they are considered solitonic as they often appear as isolated lumps of energy.

Studying symmetric examples of topological solitons has historically paved the way towards a greater understanding of these objects from both a geometric and physical perspective. For one thing, imposing symmetry is a realistic method of finding explicit, or at least semi-explicit, solutions to complicated problems. One can also gain geometric insight; for example, it is known that the fixed point set under an isometry group is always a \textit{totally geodesic submanifold} of the moduli space. This is useful for studying dynamics of such solitons, as trajectories are expected to be along geodesics. Symmetries have been explored in great detail for instantons, monopoles, and skyrmions, for example in  \cite{AllenSutcliffe2013,battyesutcliffe1997symmetricskyme,BraatenTownsendCarson1990,HitchinMantonMurray1995,MantonSutcliffe2014,NorburyRomao2007spectral,Sutcliffe1996cyclic}.

Calorons are topological solitons, which when interpreted physically describe instantons on $\R^4$ at finite temperature. In such a scenario, the euclidean time is seen to be periodic, and so calorons are often given the alternative name of \textit{periodic instantons}. Mathematically, this is encapsulated in the language of finite action anti-self-dual connections on vector bundles over $S^1\times\R^3$. For many reasons, calorons provide a link between the aforementioned solitons: it has been shown that monopoles appear as large scale limits, and instantons as large period limits of certain calorons \cite{Harland2007,Rossi1979}, and another link between calorons and monopoles is that the boundary data for a caloron consists of defining \textit{constituent monopoles}, which is mathematically understood in terms of Garland and Murray's loop group approach \cite{GarlandMurray1988}. By taking the holonomy of an instanton on $\R^4$ along a line, it is known to produce a good approximation to a skyrmion \cite{AtiyahManton1989}. When one of the directions is periodic, i.e. calorons, then such a construction has been shown to approximate Skyrme chains, also known as periodic skyrmions \cite{HarlandWard2008chains}.

These links between calorons and other solitons, and the traditional interest in symmetries, gives a reasonable motivation for studying symmetric calorons. This has been done for some specific symmetries arising from isometries of $S^1\times\R^3$ (euclidean symmetries) for various special cases of calorons \cite{Harland2007,NakamulaSawado2013cyclic,Ward2004}. What has been, until now, yet to be considered are larger classes of isometries of the moduli space of calorons which do not simply stem from euclidean symmetries. In the case where the gauge transformations take values in $U(N)$, there exists an isometry of the moduli space called \textit{the rotation map}. This acts as a gauge transformation, which is not itself periodic, but still preserves the periodicity of the caloron. Its effect is best understood in terms of the constituent monopoles -- the rotation map cyclically permutes the constituent monopoles via the outer automorphism of the loop group which `rotates' its Dynkin diagram. In the cases where the monopoles have identical charges and masses, they are indistinguishable, so such a map constitutes a $\Z_N$ action on the moduli space. Whilst it can be shown that there are no fixed points under this map alone, one may still consider fixed points of euclidean symmetries coupled with this discrete isometry.

The main result of this paper is a classification of calorons invariant under the action of cyclic groups which are coupled with the rotation map (theorem \ref{theorem_cyclic_2k_rotation}). In particular, we give an explicit formula for the solutions in a special case in terms of caloron Nahm data. The paper is briefly outlined as follows: we begin in section \ref{section_calorons} with a sufficient literature review of calorons, describing their topology, and a review of known examples. Section \ref{section_constructions} is dedicated to the different data and constructions related to calorons -- the Nahm transform, and the monad construction. In section \ref{section_symmetries} we shall discuss symmetries, that is, the various group actions on the spaces of calorons, including the elusive rotation map, and some preliminary results regarding invariant solutions. In section \ref{section_cyclic} we derive our main results about cyclic calorons in the context of monad matrix data, and finally we construct some corresponding Nahm data in section \ref{section_Nahm_calculation}, which may be read almost completely independently of section \ref{section_cyclic}.
\section{Calorons}\label{section_calorons}
In short, calorons are finite action, anti-self-dual connections $D=d+A$, on $M=S^1\times\R^3$, where the metric on $M$ is the standard flat product metric which gives the circle circumference $2\pi/\mu_0$, for some number $\mu_0>0$, that is we consider
$$S^1=\bigslant{\R}{\frac{2\pi}{\mu_0}\Z}.$$ The action in question is the \textit{Yang-Mills} action
\begin{align}\label{YM-action}
    S_{YM}:=-\int_M\tr(F\wedge\star F),
\end{align}
where $F$ is the curvature of the connection $A$, and $\star$ denotes the Hodge star operator. To be \textit{anti-self-dual} means that the curvature satisfies
\begin{align}\label{ASD}
    \star F=-F.
\end{align}
The anti-self-dual (ASD) condition (\ref{ASD}) is a natural consideration alongside the action (\ref{YM-action}) since all such connections are minima, satisfying the Euler-Lagrange equations $D\star F=0$ trivially by the Bianchi identity $DF=0$.

The most common type of calorons studied are ``$SU(N)$ calorons". What this means is that the connection is equipped to a trivial, rank $N$ hermitian vector bundle $V\to S^1\times\R^3$, which possesses a parallel volume form. We shall restrict our attention to the simplest and best-understood case -- $N=2$. The set-up and definitions are similar for general $N$, however the techniques for construction are significantly less well-understood.
\subsection{Boundary conditions, topological data, and moduli spaces}
In order to obtain finite action, strict boundary conditions need to be imposed as $r\to\infty$, where $r$ denotes the radial coordinate in $\R^3$. These determine (and are determined up to isomorphism by) a set of $SU(2)$ caloron boundary data, which manifests in the form of a tuple
$$(m_1,m_2,\nu_1,\nu_2),$$
where $m_1,m_2\in\N$, $\nu_1,\nu_2\in[0,\mu_0]$, $\nu_1+\nu_2=\mu_0$. The $m_p$ and $\nu_p$ are known respectively as the \textbf{monopole charges} and \textbf{monopole masses} of the caloron.

Our conventions for the boundary conditions were first considered by Nye \cite{Nyethesis} as sufficient for the purpose of finite action and the existence of a Nahm transform. Recent work by Cherkis and collaborators on Taub-NUT instantons \cite{CherkisLarrainHubachStern2016instantons} suggests that these conditions are in fact also necessary for finite action.
\subsubsection{Framing and boundary data}
Let $S^2_\infty$ denote the $2$-sphere at $r=\infty$ in $\R^3$. Let $W$ be a trivial rank 2 hermitian vector bundle over $\R^3$, and $W_\infty$ denote its restriction to $S^2_\infty$. Consider an $SU(2)$ connection $1$-form $a_\infty$ on $W_\infty$, and a section $\Phi_\infty\in\Gamma(W_\infty^{\mathrm{ad}})$ such that $\Phi_\infty$ has eigenvalues $\pm\imath\mu$, for $0\leq\mu\leq\mu_0/2$, and such that $a_\infty$ is diagonal with respect to the eigenbundles $L_\pm$ of $\Phi_\infty$, that is
$$a_\infty=a_+-a_-,$$
where $a_\pm$ is the unique connection $1$-form on $L_\pm$ such that $da_{\pm}=\pm\imath\frac{m}{2}\Omega$, with $\Omega$ denoting the volume form on $S^2_\infty$. We also further restrict the vector bundle $V\to S^1\times\R^3$ to be \textbf{framed} by $W_\infty$, that is there exists a bundle isomorphism, called the \textbf{framing},
\begin{align}\label{framing}f:V_\infty\equiv\left.V\right|_{S^1\times S^2_\infty}\to p^\ast_\infty W_\infty,\end{align}
where $p_\infty:S^1\times S^2_\infty\to S^2_\infty$ is the projection. $(V,f)$ is an example of a \textbf{framed bundle}. Now we may formulate a definition for a caloron in terms of these boundary conditions. We denote coordinates on $S^1\times\R^3$ by $(t,\vec{x})$.
\begin{definition}
Let $(V,f)$, $W_\infty$, $a_\infty$, and $\Phi_\infty$ be as above. A connection $A$ on $V\to S^1\times\R^3$ is an $SU(2)$\textbf{ caloron configuration} if its curvature is anti-self-dual, and if
\begin{align}\label{bdy_constraint}
    \left.A\right|_{S^1\times S^2_\infty}=(\Phi_\infty\circ p_\infty)dt+p_\infty^\ast a_\infty,
\end{align}
where $p_\infty:S^1\times S^2_\infty\to S^2_\infty$ is the projection.
\end{definition}
The monopole charges $m_p$ and monopole masses $\nu_p$ of the caloron may be identified from the previously defined boundary conditions. The masses are simply given in terms of the eigenvalues of $\Phi_\infty$, namely
$$\nu_1=2\mu,\quad\nu_2=\mu_0-2\mu.$$
The charges are given in terms of the \textbf{magnetic charge} $m$, and the \textbf{instanton number} $k$. These are both integers, and determine the monopole charges via
$$m_1=m+k,\quad m_2=k.$$
The magnetic charge is given by the first Chern number $c_1(L_+)$ of the eigenbundle $L_+$. To understand the instanton number, we refer the reader to Nye's thesis \cite{Nyethesis} for the details, but roughly speaking, $k$ is a measure of how the caloron \textit{winds} around the interior of $S^1\times\R^3$.

A gauge transformation $g:S^1\times\mathbb{R}^3\longrightarrow SU(2)$ is called \textit{framed} if it is identity at $r=\infty$. The moduli spaces $\mathcal{C}(m_1,m_2,\nu_1,\nu_2)$ are formed as the orbits under framed gauge transformations of $SU(2)$-caloron configurations, with boundary data $(m_1,m_2,\nu_1,\nu_2)$. An element of the moduli space $\mathcal{C}(m_1,m_2,\nu_1,\nu_2)$ is called a \textbf{framed }$\mathbf{(m_1,m_2)}$\textbf{-caloron}. The caloron boundary data $(m_1,m_2,\nu_1,\nu_2)$ combine together to form the total \textit{caloron charge}
\begin{align}\label{caloron_charge_constituents}
Q=\frac{1}{\mu_0}(m_1\nu_1+m_2\nu_2).
\end{align}
This is a topological quantity, in that it cannot be altered by continuous deformations of the caloron, and may be computed in terms of an integral, namely
\begin{align}\label{caloron_charge}
    Q=\frac{1}{8\pi^2}\int_{S^1\times\R^3}\tr(F\wedge F).
\end{align}
\subsection{Historical review of calorons}
In 1978, Harrington and Shepard wrote down the first examples of calorons \cite{HarringtonShepard1978periodic} by adapting the Corrigan-Fairlie-'tHooft ansatz to periodic instantons. These are seen to be $(k,k)$-calorons, but with monopole masses $(0,\mu_0)$, in other words, one monopole is mass-less. This corresponds to boundary data with $\mu=0$, and such calorons are said to have \textit{trivial holonomy}. In 1998, both Lee and Lu \cite{LeeLu1998}, and Kraan and van Baal \cite{KraanVanBaal1998}, independently studied the first examples of calorons with non-trivial holonomy. They considered $(1,1)$-calorons, and in particular, for certain calorons, the moduli space and metric was explicitly constructed, and found to given by the Taub-NUT space modulo a $\Z_2$ action \cite{KraanVanBaal1998}. Calorons with differing monopole charges are less understood, however some interesting examples may be found in \cite{chakrabarti1987periodic}. A simple case of such calorons are \textit{monopoles} on $\R^3$ \cite{Harland2007,Sutcliffe1997BPS}.

A truly remarkable property of ASD connections is the existence of a non-linear transform, known as the \textit{Nahm transform}, and this has to date been the most successful method towards understanding these objects. This construction and other related topics have been well-studied for $SU(2)$ calorons, and we describe this in more detail in section \ref{section_constructions}. This was explicitly utilised for the construction of the $(1,1)$ calorons in \cite{KraanVanBaal1998,LeeLu1998}. A few other examples of calorons have been implicitly described via the Nahm transform. Some charge $(k,k)$ cases were considered in \cite{bruckmannvanBall2002multi}, and many examples for $k=2$ have been written down explicitly \cite{BruckmannNogvanB2003constituent,NakamulaSakaguchi2010multicalorons,NogradiThesis}, some of which we comment upon in section \ref{section_Nahm_calculation}. Examples which exploit symmetry are given by Harland \cite{Harland2007}, Nakamula and Sawado \cite{NakamulaSawado2013cyclic}, and Ward \cite{Ward2004}, but only considering low charge $(k,k)$, and $(k+1,k)$ cases, and the latter two only with trivial holonomy.
\section{Constructing calorons}\label{section_constructions}
In general, finding explicit solutions to the ASD equations is a difficult problem, which makes getting a handle on the moduli spaces hard as well. In this section we shall describe two sets of objects -- Nahm data, and monad matrices -- both of which have moduli in 1-1 correspondence with calorons. These objects do not rely on solving PDEs to write them down, and so studying them provides a more user-friendly approach to understanding the moduli spaces of calorons.
\subsection{Nahm transforms}
In general, a \textit{Nahm transform} is a bijection between moduli of anti-self-dual connections with moduli of anti-self-dual connections defined on a ``dual" space \cite{jardim2004}. The ``dual" data is usually known as \textit{Nahm data}. The origins of this transform lies in the $ADHM$ construction of instantons \cite{ADHM1978construction}, and later, the original `Nahm transform' for monopoles \cite{Nahm1983trsfm}. The Nahm transform for $SU(N)$ calorons was shown to exist by Nye \cite{Nyethesis}, and finally proven to be a bijection in the case $N=2$ by Charbonneau and Hurtubise \cite{CharbonneauHurtubise2007nahm.tfm}. There are still many open problems involving Nahm transforms for calorons, for example, both the moduli space of calorons and the moduli space of Nahm data are equipped with a hyper-k\"ahler metric, but it is still only conjectured that the Nahm transform is an isometry.
\subsubsection{Caloron Nahm data}
For the purpose of this article, we are ultimately only interested in the Nahm data for $(k,k)$-calorons. Here we describe such data. Consider the circle $S^1=\bigslant{\R}{\mu_0\Z}$, covered by the intervals $I_1=[-\mu,\mu]$ and $I_2=[\mu,\mu_0-\mu]$.
\begin{definition}
Let $T_p^\lambda(s):I_p\to\mathfrak{u}(k)$ be smooth functions, and $(u_p,w_p)\in(\C^k\times\C^k)\backslash\{(0,0)\}$, for $p\in\{1,2\}$ and $\lambda\in\{0,1,2,3\}$. Define the matrix-valued functions
\begin{align*}
A_p(s)(\zeta)=T_p^2(s)+\imath T_p^3(s)+2\imath T_p^1(s)\zeta+(T_p^2(s)-\imath T_p^3(s))\zeta^2,\quad s\in I_p,\:\zeta\in\C,
\end{align*}
$p=1,2$. A set of $k$-\textbf{Nahm data} is given by a collection $(T_p^0,T_p^1,T_p^2,T_p^3,(u_p,w_p))_{p=1,2}$ such that
\begin{enumerate}
\item The matrices $T_p^\lambda$ satisfy \textit{Nahm's equations}
\begin{align}\label{Nahm_eqn}
\frac{d}{ds}T_p^j+[T_p^0,T_p^j]+\frac{1}{2}\sum_{k,l=1}^3\epsilon_{jkl}[T_p^k,T_p^l]=0,
\end{align}
for $j\in\{1,2,3\}$.
\item The matrices $A_p(s)(\zeta)$ have well-defined limits at $\bdy I_p$ in a gauge where $T_p^0=0$, and then the column vectors $(u_p,w_p)$ and the matrices $A_p(s)(\zeta)$ satisfy the matching conditions
\begin{align}\label{matching_cond}
\begin{array}{r}
A_1(\mu)(\zeta)-A_2(\mu)(\zeta)=(u_1-w_1\zeta)(w_1^\dagger+u_1^\dagger\zeta),\\
A_2(\mu_0-\mu)(\zeta)-A_1(-\mu)(\zeta)=(u_2-w_2\zeta)(w_2^\dagger+u_2^\dagger\zeta),
\end{array}
\end{align}
for all $\zeta\in\C$.
\end{enumerate}
\end{definition}
We may think of the matrices $T_p^j$, $j=1,2,3$, as endomorphisms of a rank $k$ hermitian vector bundle $K_p$ over $I_p$ with gluing conditions given by the matching equations (\ref{matching_cond}) above to make a hermitian bundle $K$ over the circle $\bigslant{\R}{\mu_0\Z}$. The matrices $T_p^0$ may be thought of as connection one-forms on $K_p$. 
\begin{definition}
A collection of $k$-Nahm data is called \textbf{irreducible} if there are no parallel sections of $K_p$ invariant under the matrices $T_p^j$ for each $p=1,2$, $j=1,2,3$.
\end{definition}
It turns out that it is precisely the irreducible Nahm data which are needed in order to perform a bijective Nahm transform. We can describe the total (irreducible) $k$-Nahm-data configuration space $\mathcal{T}_k$ as the set of tuples containing all Nahm matrices $T_p^\lambda$ for both intervals, and the corresponding matching vectors $(u_p,w_p)$. For notational brevity, we will often write the elements of $\mathcal{T}_k$ as $(T_p^\lambda,(u_p,w_p))$. There are gauge transformations which act on Nahm data given by the elements of the gauge group
$$\GG_k=\left\{(g_1,g_2)\mbox{ }|\mbox{ }g_p:I_p\to U(k),\mbox{ }g_1(\mu)=g_2(\mu),\mbox{ }g_1(-\mu)=g_2(\mu_0-\mu)\right\}$$
which, for $g=(g_1,g_2)\in\GG_k$, acts on $\mathcal{T}_k$ via $g\cdot(T_p^\lambda,(u_p,w_p))=\left(g\cdot T_p^\lambda,g\cdot (u_p,w_p)\right)$, where
\begin{align*}
\left(g\cdot T_p^j\right)(s)&=g_p(s)T_p^j(s)g_p(s)^{-1},\quad j\in\{1,2,3\},\\
\left(g\cdot T_p^0\right)(s)&= g_p(s)T_p^0(s)g_p(s)^{-1}-\frac{dg_p}{ds}(s)g_p(s)^{-1},\\
g\cdot (u_1,w_1)&= g_1(\mu)(u_1,w_1),\\
g\cdot(u_2,w_2)&=g_2(\mu_0-\mu)(u_2,w_2).
\end{align*}
The moduli space of (irreducible) $k$-Nahm data is given by $\mathcal{N}_k=\bigslant{\mathcal{T}_k}{\GG_k}$.
For details of the Nahm transform for calorons, that is, the method of obtaining a caloron from an element of $\mathcal{N}_k$, see \cite{CharbonneauHurtubise2007nahm.tfm,Nyethesis}.
\subsection{The monad construction}
The Nahm transform identifies calorons -- solutions to a system of PDEs -- with Nahm data -- solutions to a system of ODEs, hence this construction makes the writing down of explicit solutions more accessible. For calorons, it is possible to go one step further and reduce these ODEs into a system of algebraic equations. The accompanying data (monad matrices) are a by-product of the \textit{monad construction} \cite{CharbonneauHurtubise2008Rat.map}, and we aim to give a brief overview of their definitions and how they come about in this section. We shall also cover some important basic properties which will be helpful later on.
\begin{definition}
Let $A,B\in Mat_{k\times k}(\C)$, $C\in Mat_{k\times 2}(\C)$ and $D\in Mat_{2\times k}(\C)$. $(A,B,C,D)$ are called \textbf{monad matrices} if they satisfy the \textit{monad equation}
\begin{align}\label{monad_eqn}
[A,B]+CD=0,
\end{align}
and the full-rank conditions: the linear maps given by
\begin{align}\label{full-rank_conditions}
\begin{pmatrix}
A-y\mathbb{1}_k\\
B-x\mathbb{1}_k\\
D\end{pmatrix},\quad\text{and}\quad
\begin{pmatrix}
x\mathbb{1}_k-B&A-y\mathbb{1}_k&C\end{pmatrix},
\end{align}
are injective and surjective respectively, for all $x,y\in\C$. We denote the configuration space of monad matrices by $\wh{\mathrm{\textbf{Md}}}_k$. The \textbf{moduli space of monad matrices} is given by $\wh{\MM}_k=\bigslant{\wh{\mathrm{\textbf{Md}}}_k}{\sim}$, where
\begin{align}
(A,B,C,D)\sim(gAg^{-1},gBg^{-1},gC,Dg^{-1}),\label{gt_monads}
\end{align}
for some $g\in GL(k,\C)$.
\end{definition}
The notation $C_p$ and $D_p$ shall always represent the $p$-th column of $C$ and the $p$-th row of $D$ respectively.
\begin{definition}
Monad matrices such that $A\in GL(k,\C)$ are known as \textbf{non-singular monad matrices}. We denote the configuration and moduli spaces by $\mathrm{\textbf{Md}}_k$ and $\MM_k$ respectively.
\end{definition}
The underlying result regarding the construction of caloron moduli spaces is the following:
\begin{theorem}[Charbonneau \& Hurtubise, 2008 \cite{CharbonneauHurtubise2008Rat.map,CharbonneauHurtubise2007nahm.tfm}]\label{cal_nahm_monad_equiv} The following are holomorphically equivalent:
\begin{enumerate}
\item The moduli space $\mathcal{C}(k,k,\nu_1,\nu_2)$ of (framed) $(k,k)$-calorons.
\item The moduli space $\mathcal{N}_k$ of (irreducible) $k$-Nahm data.
\item The moduli space $\MM_k$ of non-singular monad matrices.
\end{enumerate}
\end{theorem}
\begin{remark}
The full moduli space of monad matrices $\wh{\MM}_k$ parameterises the moduli space of instantons on $S^4$ \cite{donaldson1984instantons}. The condition of non-singularity is specific to calorons, and provides an embedding of caloron moduli spaces into instanton moduli spaces.
\end{remark}
The equivalence between 1 and 2 is given by the Nahm transform for calorons \cite{CharbonneauHurtubise2007nahm.tfm}. The equivalence between 2 and 3 is described under a Kobayashi-Hitchin correspondence, by considering a holomorphic version of the Nahm data, called a \textbf{Nahm complex}. A Nahm complex $(\alpha,\beta,u,w)$ is obtained from a set of $k$-Nahm data via
\begin{align*}
    \alpha_p(s)=T_p^0(s)+\imath T_p^1(s),\quad\beta_p(s)=T_p^2(s)+\imath T_p^3(s),\quad u=(u_1,u_2),\quad w=(w_1,w_2).
\end{align*}
Nahm's equations are hence equivalent to the \textit{complex} and \textit{real} equations
\begin{align}
\frac{d\beta_p}{ds}(s)+[\alpha_p(s),\beta_p(s)]&=0,\label{complex_eqn_cal}\\
\frac{d}{ds}\left(\alpha_p(s)+\alpha_p(s)^{\dagger}\right)+[\alpha_p(s),\alpha_p(s)^\dagger]&+[\beta_p(s),\beta_p(s)^\dagger]=0.\label{real_eqn_cal}
\end{align}
A theorem of Donaldson \cite{donaldson1984} in the context of $SU(2)$ monopoles says that the orbits of solutions to (\ref{complex_eqn_cal}) under the action of complex gauge transformations $g_p:I_p\to GL(k,\C)$ contain a unique solution to both (\ref{complex_eqn_cal}) and (\ref{real_eqn_cal}). This result may easily be extended to calorons \cite{CharbonneauHurtubise2008Rat.map}. 
Given a Nahm complex $(\alpha,\beta,u,w)$, we may obtain non-singular monad matrices by setting for example
\begin{align}\label{monads_from_NC}
\begin{array}{ll}
A=\Omega(\mu_0,0)^{-1},&B=\beta_1(0),\\
C_1=\Omega(\mu,0)^{-1}u_1,&C_2=\Omega(\mu_0-\mu,0)^{-1}u_2,\\
D_1=w_1^\dagger\Omega(\mu_0,\mu)^{-1},&D_2=w_2^\dagger\Omega(\mu_0,\mu_0-\mu)^{-1},
\end{array}
\end{align}
where $\Omega$ is the parallel transport operator with respect to the connection $\alpha$. The gauge transformations acting on these matrices as constructed are given by $g=g_1(0)$. It is a straight-forward exercise to see that the complex Nahm equation, and matching conditions imply the monad equations. The full-rank conditions come from the irreducibility of the Nahm data. For more details of this, and to see that the correspondence is invertible, we direct the reader to \cite{CharbonneauHurtubise2008Rat.map}.
\subsubsection{Basic properties of monad matrices}
Here we cover some basic properties of monad matrices which shall be utilised later on. Most of these results may be found in \cite{FurutaHashimoto1990}, but we have reformulated them for our own purposes. The notation we use for monad matrices may be translated to that in \cite{FurutaHashimoto1990} via $(A,B,C,D)\leftrightarrow(\alpha_1,\alpha_2,b,a)$.

Let $\II$ denote the set of all words $\Pi(x_1,x_2)$ in two variables $x_1$ and $x_2$.
\begin{lemma}\label{words_cond} $(A,B,C,D)\in\wh{\mathrm{\textbf{Md}}}_k$ if and only if $[A,B]+CD=0$ and the following conditions hold:
\begin{flalign}\label{words_cond_1}
\text{For all }v\in\C^k\setminus{0}\text{, there exists }\Pi\in\II\text{ such that }D\Pi(A,B)v\neq 0.&
\end{flalign}
\begin{flalign}\label{words_cond_2}
\sum_{\Pi\in\II}\Pi(A,B)\Img(C)=\C^k.
\end{flalign}
\end{lemma}
\textit{Proof}. See \cite{FurutaHashimoto1990}.\hfill$\square$
\begin{corollary}\label{C,Dnon0}
Let $(A,B,C,D)\in\wh{\MM}_k$. Then $C,D\neq 0$.
\end{corollary}
\textit{Proof}. If $C=0$ or $D=0$, this violates (\ref{words_cond_2}) or (\ref{words_cond_1}) respectively.\hfill$\square$
\begin{lemma}\label{free_gauge_monads}
The action of gauge transformations (\ref{gt_monads}) is free on $\wh{\mathrm{\textbf{Md}}}_k$.
\end{lemma}
\textit{Proof}. Let $(A,B,C,D)\in\wh{\mathrm{\textbf{Md}}}_k$ and suppose that $g\in GL(k,\C)$ is such that
$$(A,B,C,D)=(gAg^{-1},gBg^{-1},gC,Dg^{-1}).$$
Let $U=\Img(\mathbb{1}-g)$, and $u=(\mathbb{1}-g)v\in U$. Then for all $\Pi\in\II$, we have
$$D\Pi(A,B)u=D\left(\Pi(A,B)v-g\Pi(A,B)g^{-1}gv\right)=D\Pi(A,B)v-Dg^{-1}g\Pi(A,B)v=0.$$
Hence, by lemma \ref{words_cond} condition (\ref{words_cond_1}), we must have $U=0$, i.e. $g=\mathbb{1}$.\hfill$\square$
\begin{lemma}\label{lemma.simul_diag}
Let $(A,B,C,D)\in\wh{\MM}_k$. Then the matrices
$$\{D\Pi(A,B)C\::\:\Pi\in\II\}$$
cannot be made simultaneously diagonal.
\end{lemma}
\textit{Proof}. See \cite{FurutaHashimoto1990}.\hfill$\square$
\section{Symmetric calorons}\label{section_symmetries}
In this section, we shall describe different group actions on the moduli spaces $\mathcal{C}(m_1,m_2,\nu_1,\nu_2)$ of $SU(2)$ calorons, and the accompanying Nahm, and monad matrix data. Ultimately, we shall be interested in studying the fixed points under these group actions. The moduli space $\mathcal{C}(m_1,m_2,\nu_1,\nu_2)$ possesses a natural riemannian metric, and we consider the group actions on $\mathcal{C}(m_1,m_2,\nu_1,\nu_2)$ which are isometries of this metric. These isometries form a group
$$\mathcal{S}\cong(O(2)\times O(3))_+\times(\Z_2\rtimes U(1)),$$
made up of the orientation preserving isometries of $S^1\times\R^3$, a \textit{phase action}, and a \textit{rotation map}, where we have omitted the inclusion of translations in $\R^3$ as clearly there are no fixed points for this.

The group $(O(2)\times O(3))_+$ acts on calorons in the natural way, via pull-back. We shall often refer to these isometries as the \textit{euclidean symmetries}, and for the duration of this article, we restrict our attention to those in subgroups of $SO(2)\times SO(3)$. Traditionally, euclidean symmetries have been the only considerations when looking for fixed points. In the next section, we shall describe further isometries -- the phase and rotation map -- which we shall utilise to find symmetric calorons.
\subsection{Phases and the rotation map}
Recall that calorons are equivalent up to the action of framed gauge transformations. A gauge transformation is \textit{unframed} if it is not identity at spatial infinity. These are isometries of calorons since the metric is gauge invariant. One type of unframed gauge transformation is a \textbf{phase}. For $\omega\in(-\pi,\pi)$, the phase action is given by a bundle map $P(\omega):V\to V$ such that its restriction to $W_\infty=L_+\oplus L_-$ is
\begin{align}\begin{array}{rcccc}
P(\omega)_\infty:&W_\infty&\to&W_\infty&\\
&p&\mapsto&e^{\pm\imath\omega}p,&\text{on the eigenbundle }L_\pm.
\end{array}
\end{align}
Another type are the \textit{large gauge transformations}, that is, $g:\R\times\R^3\to SU(2)$ such that for all $\vec{x}\in\R^3$,
\begin{align}\label{large_gt}
g(t+2\pi/\mu_0,\vec{x})=- g(t,\vec{x}).\end{align}
While these do not come from automorphisms of $V$, they still preserve the periodicity of calorons. The example of such a transformation that we are interested in is \textbf{the rotation map}, which is an identification of moduli spaces which permutes the masses and charges:
\begin{align}\label{rotation_map_cal}
    \rho_\CC:\mathcal{C}(m_1,m_2,\nu_1,\nu_2)\to\CC(m_2,m_1,\nu_2,\nu_1).
\end{align}
The map is defined as follows. For each $t\in\R$, define the family of bundle maps
\begin{align}\label{rotation_map_bdy}
    \begin{array}{rcccc}
        (\wt{\rho}_t)_\infty:&W_\infty&\to&W_\infty,&  \\
         &p&\mapsto&e^{\pm\imath t\frac{\mu_0}{2}}p,&\text{on }L_\pm.
    \end{array}
\end{align}
These may each be extended smoothly, and arbitrarily, to cover all of the interior of $\R^3$ to maps $\wt{\rho}_t:W\to W$. These in turn induce gauge transformations $\rho_t:\R^3\to SU(2)$. The rotation map is hence given by the large gauge transformation $\rho(t,\vec{x})=\rho_t(\vec{x})$, which one may easily see satisfies (\ref{large_gt}). One may also straight-forwardly see that $\rho$ provides the correspondence (\ref{rotation_map_cal}) by considering the boundary conditions for calorons \cite{Nyethesis}.

The upshot is, whilst the euclidean symmetries and phases are well-defined \textit{actions} on the moduli space $\mathcal{C}(m_1,m_2,\nu_1,\nu_2)$, the rotation map $\rho$ only ever defines an action on a moduli space in the special cases where $m_1=m_2=k$ and $\nu_1=\nu_2=\frac{\mu_0}{2}$. In these cases, it contributes an overall $\Z_2$ action on the moduli space. For this reason, we shall henceforth focus our attention to these cases. We shall also make use of the simpler notation $\CC_k$ to denote the moduli space $\mathcal{C}(k,k,\mu_0/2,\mu_0/2)$ of $(k,k)$-calorons with equal monopole masses.
\subsection{Group actions on Nahm data and monads}
We have demonstrated different actions on the moduli spaces of calorons, and in this section we summarise the corresponding actions on the Nahm and monad matrix data in the tables below. Importantly, the construction of the matrix data requires the fixing of an axis in $\R^3$, which we have chosen to be the $x^1$ axis. This is what reduces the structure to that of a holomorphic structure, and hence only a subset of the actions of $\mathcal{S}$ are compatible with this holomorphic structure, and are the only ones that are able to be straight-forwardly written down. The restrictions are imposed on the euclidean contributions 
$SO(2)\times SO(3)$, in that we may only consider spatial rotations $R_\theta\in SO(3)$ of angle $\theta$ about the fixed axis $\vec{n}_1\in\R^3$, and translations $S_\varphi\in SO(2)$ in the circle direction. We denote the components of $R$ by $R_{jk}$.
\subsubsection*{Actions on Nahm data}
\begin{center}
\begin{tabular}{c|c}
    Element of $\mathcal{S}$ & Action on $\mathcal{N}_k$ via $(T_p^\lambda,(u_p,w_p))\mapsto \star$ \\
    \hline
    $R_\theta$ & $\left(T_p^0,\sum_{k=1}^3R_{jk}T_p^k,\left(e^{-\imath\frac{\theta}{2}}u_p,e^{\imath\frac{\theta}{2}}w_p\right)\right)$\\
    $S_\varphi$ & $\left(T_p^0+\imath\frac{\varphi}{\mu_0}\mathbb{1},T_p^j,(u_p,w_p)\right)$\\
    $P(\omega)$ & $(T_p^\lambda,e^{\imath\omega}(u_1,w_1),e^{-\imath\omega}(u_2,w_2))$\\
    $\rho$ & $\left(T_2^\lambda(s+\mu_0/2),T_1^\lambda(s-\mu_0/2),(u_2,w_2),(u_1,w_1)\right)$
\end{tabular}
\end{center}
\subsubsection*{Actions on monad matrices}
\begin{center}
\begin{tabular}{c|c}
    Element of $\mathcal{S}$ & Action on $\mathcal{M}_k$ via $(A,B,C,D)\mapsto \star$ \\
    \hline
    $R_\theta$ & $\left(A,e^{-\imath\theta}B,e^{-\imath\frac{\theta}{2}}C,e^{-\imath\frac{\theta}{2}}D\right)$\\
    $S_\varphi$ & $(e^{-\imath\varphi}A,B,C\adj(\sigma_\varphi),\sigma_{\varphi} D)$\\
    $P(\omega)$ & $(A,B,Cq_\omega^{-1},q_\omega D)$\\
    $\rho$ & $(A,B-C_1D_1A^{-1},C_2,AC_1,D_2,D_1A^{-1})$
\end{tabular}
\end{center}
Here we have introduced the following short-hands:
$$\sigma_\varphi=\begin{pmatrix}
e^{-\imath\frac{3\varphi}{4}}&0\\
0&e^{-\imath\frac{\varphi}{4}}\end{pmatrix},\quad\text{and}\quad q_\omega=\begin{pmatrix}
e^{-\imath\omega}&0\\
0&e^{\imath\omega}
\end{pmatrix},$$
with $\mathrm{adj}(\sigma_{\varphi})=\det(\sigma_\varphi)\sigma_\varphi^{-1}$. We remark that it is straight-forward to write down full the action of $SO(3)$ on Nahm data, but understanding how to define this for the matrix data and similar objects remains an important open problem.
\subsection{Studying invariant solutions}
We are interested in determining various fixed points under the action of subgroups of $\mathcal{S}$:

\begin{definition}
Let $H\subset\mathcal{S}$ be a subgroup. A $(k,k)$-caloron is said to be $H$\textbf{-symmetric} if for all $h\in H$, $[h\cdot A]=[A]$. We denote the fixed point set of all $H$-symmetric calorons by $\CC_k(H)$.
\end{definition}
Using both the Nahm transform, and the construction (\ref{monads_from_NC}), one can check that the actions defined on the Nahm and monad matrix data may be identified (modulo gauge transformations) with the corresponding actions on calorons. Hence, by theorem \ref{cal_nahm_monad_equiv}, the fixed point sets of these actions (on Nahm data or on monad matrices), modulo gauge transformations, are in biholomorphic correspondence with the fixed point sets of calorons.

In sections \ref{section_cyclic} and \ref{section_Nahm_calculation}, we shall explicitly construct invariant solutions for various cyclic subgroups of $\mathcal{S}$ in the context of monad matrix data, and Nahm data. However, before we do that, it is worthwhile discussing some preliminary results regarding which of the actions may have fixed points, and which do not.
\begin{proposition}\label{rotation_phase_no_fps}
Let $H\subset\mathcal{S}$ be such that $\varphi(\omega)\rho\in H$, for some $\omega\in(-\pi,\pi)$. Then $\CC_k(H)=\emptyset$.
\end{proposition}
\textit{Proof}. Let $g_\omega=\varphi(\omega)\rho$, and suppose that $A$ is a caloron such that $[h\cdot A]=[A]$. Then
$$d(hg_\omega)+Ahg_\omega-hg_\omega A=0,$$
which implies $\frac{\bdy}{\bdy x^\lambda}\tr(hg_\omega)=0$ for all $\lambda=0,1,2,3$ (here, we identify $t\equiv x^0$). Since $\det(hg_\omega)=1$, this implies that the eigenvalues of $hg_\omega$ must be constant over $S^1\times\R^3$. In particular, they must be constant on $S^1\times S^2_\infty$, in contradiction to the form of the rotation map at $\infty$ (\ref{rotation_map_bdy}).\hfill$\square$
\begin{proposition}\label{lemma.phase_no_fp}
There are no fixed points in $\MM_k$ under the action of a phase $P(\omega)$, where $\omega\in(-\pi,\pi)\setminus\{0\}$. 
\end{proposition}
\textit{Proof}. If $(A,B,C,D)\in\mathrm{\textbf{Md}}_k$ is such that
$$(A,B,C,D)=(hAh^{-1},hBh^{-1},hCq_\omega^{-1},q_\omega Dh^{-1})$$
for some $h\in GL(k,\C)$, then for all $\Pi\in\II$,
$$D\Pi(A,B)C=q_\omega D\Pi(A,B)Cq_\omega^{-1}.$$
As $q_\omega\neq\pm\mathbb{1}$, we must have that the matrices $\{D\Pi(A,B)C\::\:\Pi\in\II\}$ are simultaneously diagonal, which is impossible by lemma \ref{lemma.simul_diag}.\hfill$\square$\\

\noindent These two results show that unframed gauge transformations alone cannot have any fixed points, and so the inclusion of non-trivial euclidean actions is imperative to find invariant solutions. Importantly, not just any euclidean symmetries can be exploited. Indeed, it is straight-forward to see that any fixed points under an action of $S^1\times S^1$, the eigenvalues of the matrices $A$ and $B$ must be preserved. This observation in particular tells us what subgroups may be considered:
\begin{lemma}\label{lemma_j_even}
Let $H\subset\mathcal{S}$ contain a generator $S_\varphi$ of translations of the circle. Then $\mathcal{C}_k(H)\neq\emptyset$ only if $\varphi=2r\pi/k$, for some $r\in\Z$.
\end{lemma}
\textit{Proof}. The action of $H$ on monad matrices affects the matrix $A\in GL(k,\C)$ via $A\mapsto e^{-\imath\varphi}A$. This is invariant in $\mathcal{M}_k$ only if the eigenvalues of $A$ are preserved. In particular, $\det A\neq 0$, so we must have
$$e^{-\imath k\varphi}=1,$$
that is, there exists $r\in\Z$ such that $\varphi=2r\pi/k$.\hfill$\square$\\

\noindent What lemma \ref{lemma_j_even} says is that the only translations of the circle that may yield fixed points are those of finite cyclic type. However, for invariant solutions, these euclidean actions must be considered along with rotations of $\R^3$. Indeed, if a $(k,k)$-caloron were seen to be invariant under the action $S_{2r\pi/k}$ alone, then this is the same as saying it has $r/k$ times its expected period. Such a caloron is more efficiently described by a lower charge caloron.
\section{Cyclic calorons}\label{section_cyclic}
The work of Braden and Sutcliffe on cyclic monopoles \cite{braden2011,Sutcliffe1996cyclic} was very influential in furthering the understanding of monopole moduli spaces, so there is hence much benefit in studying cyclic calorons for similar purposes. For calorons, we may consider euclidean cyclic subgroups of $S^1\times S^1$, that is, ones which act temporally as well as in space. To this end, let $K_m^j\in(O(2)\times O(3))_+$ be defined by
$$K_m^j:=\left(\begin{pmatrix}
\cos\left(\frac{2j\pi}{m}\right)&-\sin\left(\frac{2j\pi}{m}\right)\\
\sin\left(\frac{2j\pi}{m}\right)&\cos\left(\frac{2j\pi}{m}\right)
\end{pmatrix},\begin{pmatrix}1&0&0\\
0&\cos\left(\frac{2\pi}{m}\right)&\sin\left(\frac{2\pi}{m}\right)\\
0&-\sin\left(\frac{2\pi}{m}\right)&\cos\left(\frac{2\pi}{m}\right)
\end{pmatrix}\right),$$
for $m\in\Z^+$ and $j\in\Z_m$. This generates an order $m$ cyclic subgroup of $SO(2)\times SO(3)$. From this, we may form the following cyclic subgroups of $\mathcal{S}$: the \textbf{regular cyclic groups}
\begin{align}
C_{n}^{j,q}:=\left\langle K_{n}^j\circ P({2q\pi/n})\right\rangle\subset\mathcal{S},
\end{align}
for $n\in\Z^+$, $j,q\in\Z_{n}$, and the \textbf{rotation cyclic groups}
\begin{align}
\rho(C_{2n}^{j,\omega}):=\left\langle K_{2n}^j\circ P(\omega)\circ\rho\right\rangle\subset\mathcal{S},
\end{align}
for $n\in\Z^+$, $j\in\Z_{2n}$, $\omega\in(-\pi,\pi)$. It is straight-forward to see that $C_n^{j,q}$ and $\rho(C_{2n}^{j,\omega})$ are cyclic groups of order $n$ and $2n$ respectively, for all $j,q,\omega$. Furthermore, our analysis in the previous section shows that these are the most generic cyclic groups (up to isomorphism) which can yield fixed points in the moduli space of $(k,k)$-calorons, in particular, without repeating oneself by considering the same group in a different presentation, or considering lower charged calorons as higher charge ones.

The purpose of this article is to investigate isometries involving the rotation map, so for this reason we shall focus our attention on the fixed point set $\CC_k(\rho(C_{2k}^{j,\omega}))$. We remark that many of the same methods may be used to consider the more general cases of these groups for order $2n$, for example when $n$ divides $k$, with greater complexity as you lower the order. Our main result is the following:
\begin{theorem}\label{theorem_cyclic_2k_rotation} Let $\omega\in(-\pi,\pi)$ and $j\in\Z_{2k}$.
\begin{enumerate}
    \item If $j/2\notin\Z_k$, then $\CC_k(\rho(C_{2k}^{j,\omega}))=\emptyset$.
    \item If $j/2\in\Z_k$, then $\CC_k(\rho(C_{2k}^{j,\omega}))\cong(\C^\ast)^2$.
\end{enumerate}
\end{theorem}
What follows in this section is a proof of this theorem. It is, however, worth mentioning that these rotation cyclic groups are the only collection of cyclic isometries which remain unstudied in the literature. Special cases of $C_k^{0,q}$-invariant calorons were studied in some detail in \cite{NakamulaSawado2013cyclic}, but more generally, $C_n^{j,q}$-invariant instantons were fully classified by Furuta and Hashimoto in \cite{FurutaHashimoto1990} by utilising the monad matrices, and hence the corresponding cyclic calorons may be understood implicitly via the embedding $\MM_k\subset\wh{\MM}_k$.

The set $\CC_k(\rho(C_{2k}^{j,\omega}))$ is parameterised by $\rho(C_{2k}^{j,\omega})$-invariant non-singular monad matrices, that is, $(A,B,C,D)\in\MM_k$ such that
\begin{align}
e^{\imath\frac{j\pi}{k}}A&=gAg^{-1},\label{C_2k_rot_A}\\
e^{\imath\frac{\pi}{k}}B&=g\left(B-C_1D_1A^{-1}\right)g^{-1},\label{C_2k_rot_B}\\
e^{\imath\frac{\pi}{2k}}e^{\imath\frac{j\pi}{4k}}C_1&=e^{\imath\omega}gC_2,\label{C_2k_rot_C_1}\\
e^{\imath\frac{\pi}{2k}}e^{\imath\frac{3j\pi}{4k}}C_2&=e^{-\imath\omega}gAC_1,\label{C_2k_rot_C_2}\\
e^{\imath\frac{\pi}{2k}}e^{\imath\frac{3j\pi}{4k}}D_1&=e^{-\imath\omega}D_2g^{-1},\label{C_2k_rot_D_1}\\
e^{\imath\frac{\pi}{2k}}e^{\imath\frac{j\pi}{4k}}D_2&=e^{\imath\omega}D_1A^{-1}g^{-1},\label{C_2k_rot_D_2}
\end{align}
for some $g=g_{j,\omega}\in GL(k,\C)$. We shall consider also $C_k^{j,0}$-invariant matrices:
\begin{align}
e^{\imath\frac{2j\pi}{k}}A&=GAG^{-1},\label{C_k_A}\\
e^{\imath\frac{2\pi}{k}}B&=GBG^{-1},\label{C_k_B}\\
e^{\imath\frac{\pi}{k}}C&=GC\adj(\sigma_{2j\pi/k}),\label{C_k_C}\\
e^{\imath\frac{\pi}{k}}D&=\sigma_{2j\pi/k}DG^{-1},\label{C_k_D}
\end{align}
for $G\in GL(k,\C)$. We denote by $\MM_k^j(\rho^2)$ the set of matrices in $\MM_k$ satisfying (\ref{C_k_A})-(\ref{C_k_D}), and denote by $\MM_k^j\left(\rho\right)$ the set of matrices in $\MM_k^j(\rho^2)$ satisfying (\ref{C_2k_rot_A})-(\ref{C_2k_rot_D_2}). The purpose behind the second consideration is due to the inclusion $\MM_k^j(\rho)\subset\MM_k^j(\rho^2)$, which is seen by setting
\begin{align}\label{g_G_convert}
G=e^{-\imath\frac{j\pi}{2k}}g^2A.
\end{align}
Solving these equations modulo the action of $GL(k,\C)$ is all that is needed to understand the space $\CC_k(\rho(C_{2k}^{j,\omega}))$. The route of the proof of theorem \ref{theorem_cyclic_2k_rotation} is via a series of lemmas which we summarise here:
\begin{itemize}
    \item First we show that we may fix a gauge such that the matrix $G$ appearing in (\ref{C_k_A})-(\ref{C_k_D}) is a diagonal matrix, which we call the \textit{standard form}.
    \item We then show that $\det B\neq 0$, which means we may further fix a gauge so that $B$ is parameterised by its determinant.
    \item Finally we fix the matrix $g$ appearing in (\ref{C_2k_rot_A})-(\ref{C_2k_rot_D_2}) in such a way which preserves these gauge choices, and this allows us to solve the equations completely.
\end{itemize}
Note that part 1 of the theorem follows immediately from lemma \ref{lemma_j_even}, so from now on we shall assume $j$ is even.
\begin{notation}
We shall always use the following denotations: $\aleph=\mathrm{gcd}(j,k)$, $\Omega_j=e^{\imath\frac{(2-j)\pi}{2k}}$, and $\omega_r=e^{\imath\frac{2\pi r}{k}}$. Also, all indexed objects are understood to obey their respective modular arithmetic.
\end{notation}
\begin{lemma}\label{lemma_G_diag}
We may fix a basis for the matrices in $\MM_k^j(\rho^2)$ such that $$G=\Omega_j\diag\{x_1,\dots,x_k\},$$ where $x_r$, $r=1,\dots,k$, are $k$-th roots of unity.
\end{lemma}
\textit{Proof}. If $(A,B,C,D)\in\MM_k^j(\rho)$, then we may apply the symmetry equations (\ref{C_k_A})-(\ref{C_k_D}) $k$ times to yield
\begin{align*}
\begin{array}{cccc}
    A=G^kAG^{-k},&
    B=G^kBG^{-k},&
    C=-G^kC\adj(\sigma_{2j\pi}),&
    D=-\sigma_{2j\pi}DG^{-k}.
    \end{array}
\end{align*}
Since $j$ is even $\sigma_{2j\pi}=e^{\imath\frac{j\pi}{2}}\mathbb{1}$. By lemma \ref{free_gauge_monads}, the above equations imply that $-e^{\imath\frac{j\pi}{2}}G^k=\mathbb{1}$. So $G$ is diagonalisable with eigenvalues $\lambda$ satisfying $\lambda^k=e^{\imath\pi\frac{2-j}{2}}$.\hfill$\square$
\begin{definition}
$G$ is said to be of \textbf{standard form} if it is as in lemma \ref{lemma_G_diag} with $x_r=\omega_r\equiv e^{\imath\frac{2r\pi}{k}}$, $r=1,\dots,k$.
\end{definition}
Note that the order of the eigenvalues of $G$ is not important, as we may perform a gauge transformation in the form of a permutation matrix to change it. With this in mind, we may refer to any $G$ where there are $k$ distinct eigenvalues $\Omega_j\omega_r$, each of multiplicity $1$, to be of standard form. The culmination of all the next lemmas is lemma \ref{lemma_G_std}, where we prove that $G$ must be of standard form.
\begin{lemma}\label{lemma_C,D_evec}
Let $(A,B,C,D)\in\MM_k^j(\rho)$. Then $C_p,D_p\neq 0$ for all $p=1,2$. In particular,
$$\left\{\Omega_j\omega_r\::\:r=\frac{j}{2},j,k-\frac{j}{2}-1,k-1\right\}\subset\mathrm{Eval}(G),$$
with
\begin{align*}
\begin{array}{c}
     C_1\in\mathrm{Evec}\left(G,\Omega_j\omega_{\frac{j}{2}}\right),\quad
     C_2\in\mathrm{Evec}\left(G,\Omega_j\omega_j\right),\\
     D_1\in\mathrm{Ecovec}\left(G,\Omega_j\omega_{k-\frac{j}{2}-1}\right),\quad D_2\in\mathrm{Ecovec}\left(G,\Omega_j\omega_{k-1}\right).
\end{array}
 \end{align*}
\end{lemma}
\textit{Proof}. If $C_p=0$ for some $p\in\{1,2\}$, then by (\ref{C_2k_rot_C_1}) and (\ref{C_2k_rot_C_2}), this implies $C_q=0$ for $q=p+1\mbox{ }(\text{mod }2)$, i.e. $C=0$. But this violates corollary \ref{C,Dnon0}. A similar argument holds for $D_p$. The statement about eigenvectors follows from equations (\ref{C_k_C}) and (\ref{C_k_D}).\hfill$\square$
\begin{lemma}\label{lemma_evals_G}
If $\omega_i\in\mathrm{Eval}(G/\Omega_j)$ with multiplicity $m$, then $\omega_{r+\aleph p}\in\mathrm{Eval}(G/\Omega_j)$ with multiplicity $m$, for all $p=1,\dots,k/\aleph$.
\end{lemma}
\textit{Proof}. Let $\wt{G}=G/\Omega_j$, and $\{v_i\}_{i=1}^m\subset\C^k$ be a linearly independent set of vectors such that $\wt{G}v_i=\omega_rv_i$, for all $r=1,\dots,m$. By (\ref{C_k_A}), we hence have
$$\wt{G}A^{p}v_i=(\wt{G}A\wt{G}^{-1})^p\wt{G}v_i=\omega_{r+jp}A^pv_i.$$
As $A\in GL(k,\C)$, $A^pv_i\neq 0$ for all $p$, so are eigenvectors, moreover, $A^pv_i\neq A^pv_{i'}$ for all $i\neq i'$, so the eigenvalues $\omega_{r+jp}$ each have multiplicity $m$. The eigenvalues are all distinct for all $p=1,\dots,k/\aleph$, moreover
$$\{\omega_{r+jp}\::\:p=1,\dots,k/\aleph\}=\{\omega_{r+\aleph p}\::\:p=1,\dots,k/\aleph\}.$$\hfill$\square$\\

\noindent We may now prove the first main lemma in the proof.
\begin{lemma}\label{lemma_G_std} We may fix a basis for $\MM_k^j(\rho)$ such that $G$ is of standard form.
\end{lemma}
\textit{Proof}. Invoking lemma \ref{lemma_G_diag}, let $X=\mathrm{Eval}(G/\Omega_j)\subset\{\omega_r\::\:r=1,\dots,k\}$. If $\omega_r\in X$, let $V_r$ denote the eigenspace of $G/\Omega_j$ with eigenvalue $\omega_r$, and otherwise set $V_r=0$. We may form the corresponding multiplicity eigenspaces $W_p$, $p\in\{1,\dots,\aleph\}$ as
$$W_p:=\bigoplus_{r=p\mbox{ }(\text{mod }\aleph)}V_r.$$
Note by lemma \ref{lemma_evals_G}, $G$ is standard if and only if $W_p\neq 0$ for all $p$. Also by (\ref{C_k_A}) (as in the proof of lemma \ref{lemma_evals_G}), $A(W_p)\subset W_p$ for all $p=1,\dots,\aleph$. We also have by (\ref{C_k_B}) that $B(W_p)\subset W_{p+1}$ for all $p=1,\dots,\aleph$.
\paragraph{Case 1: $j/2=\aleph/2\mbox{ }(\text{mod }\aleph)$.} By lemma \ref{lemma_C,D_evec}, $W_{\frac{\aleph}{2}-1},W_{\frac{\aleph}{2}},W_{\aleph-1},W_\aleph\neq 0$, so by lemma \ref{lemma_evals_G}, for $\aleph<4$, $G$ is standard form, so we may assume $\aleph\geq 4$. Suppose that $G$ is non-standard form. By lemma \ref{lemma_evals_G}, this means there exists $1\leq r\leq \aleph-2$, $r\neq \frac{\aleph}{2}-1,\frac{\aleph}{2}\mbox{ }(\text{mod }\aleph)$ such that $W_r=0$. Hence we may either choose $\frac{\aleph}{2}\leq q_1\leq \aleph-3$, and $q_1+1< q_2\leq \aleph-1$, such that $\omega_{q_1},\omega_{q_2}\in X$, and $W_p=0$ for all $q_1<p<q_2$, or $0\leq q_3\leq \frac{\aleph}{2}-3$ and $q_3+1<q_4\leq\frac{\aleph}{2}-1$, such that $\omega_{q_3},\omega_{q_4}\in X$, and $W_p=0$ for all $q_3<p<q_4$, where these inequalities are understood modulo $\aleph$. Then, for $p=1,3$, by (\ref{C_k_B}) and lemma \ref{lemma_C,D_evec} we have
$$B(W_{q_p})=D(W_{q_p})=0.$$
Hence, as $A$ fixes each space $W_p$, for all $w\in W_{q_p}$, $p=1,3$, and for all $\Pi\in\II$, we have
$$D\Pi(A,B)w=0,$$
which violates lemma \ref{words_cond}.
\paragraph{Case 2: $j/2=0\mbox{ }(\text{mod }\aleph)$.} By lemma \ref{lemma_C,D_evec}, we have $W_{\aleph-1},W_\aleph\neq0$, so by lemma \ref{lemma_evals_G}, for $\aleph<3$, $G$ is standard form, so we may assume $\aleph\geq 3$. The argument then follows in a similar and simpler vein to case 1, leading to a violation of lemma \ref{words_cond} if $G$ is not of standard form. \hfill$\square$\\

\noindent Let $\vec{e}_i$ and $\vec{f}_i$, $i=1,\dots,k$ denote the standard basis vectors and covectors for $\C^k$ and $(\C^\dagger)^k$, and let $S$ denote the \textbf{standard shift matrix}
$$S=\left(\begin{array}{c|c}
  \begin{array}{ccc}
  0&\cdots&0\end{array} & 1 \\ 
  \hline
  \mathbb{1}_{k-1} &\begin{array}{c} 0\\\vdots\\0\end{array}
 \end{array}\right).$$
Since, by lemma \ref{lemma_G_std}, $G$ is necessarily of standard form, we may solve (\ref{C_k_A})-(\ref{C_k_D}) and hence write an ansatz for a representative of $\MM_k^j(\rho)$ in terms of these objects:
\begin{align}\label{ABCD_C_k}
\begin{array}{cc}
    A=\mathrm{diag}\{\alpha_1,\dots,\alpha_k\}S^j & B=\mathrm{diag}\{\beta_1,\dots,\beta_k\}S,\\
    \begin{array}{cc}
    C_1=u\vec{e}_{\frac{j}{2}}, & C_2=v\vec{e}_j,
    \end{array} & \begin{array}{cc}
    D_1=y\vec{f}_{k-\frac{j}{2}-1}, & D_2=z\vec{f}_{k-1}.
    \end{array}
\end{array}
\end{align}
\begin{remark}\label{ABCD,std_basis_remark}
Consider $(A,B,C,D)$ as in (\ref{ABCD_C_k}). Then we have
\begin{align}
A^m\vec{e}_p&=\left(\prod_{r=1}^m\alpha_{p+rj}\right)\vec{e}_{p+mj},\label{C_k_A_on_basis}\\
B^m\vec{e}_p&=\left(\prod_{r=1}^m\beta_{p+r}\right)\vec{e}_{p+m},\label{C_k_B_on_basis}\\
\Img(C)&=\mathrm{sp}_{\C^\ast}\{e_\frac{j}{2},e_j\},\label{C_k_C_on_basis}\\
D\vec{e}_p&=y\delta_{p,k-\frac{j}{2}-1}\begin{pmatrix}
1\\0
\end{pmatrix}+z\delta_{p,k-1}\begin{pmatrix}
0\\1
\end{pmatrix},\label{C_k_D_on_basis}
\end{align}
for all $m\in\Z^+$.
\end{remark}
The parameters $\alpha_p,u,v,y,z\in\C^\ast$, and $\beta_p\in\C$ are ultimately constrained by the monad equation (\ref{monad_eqn}), full-rank-conditions (\ref{full-rank_conditions}), and the symmetry equations (\ref{C_2k_rot_A})-(\ref{C_2k_rot_D_2}). We would like to fix a gauge such that the symmetry equations are more manageable. To do this carefully, we may make use of our next important lemma.
\begin{lemma}\label{lemma_detBnon0}
$(A,B,C,D)\in\MM_k^j(\rho)$ only if $\det B\neq 0$.
\end{lemma}
\textit{Proof}. By lemma \ref{lemma_G_std}, we may write $(A,B,C,D)\in\MM_k^{j}(\rho)$ in the form as in (\ref{ABCD_C_k}). With this data, for $j\neq 2k$, the monad equations (\ref{monad_eqn}) take the form
\begin{align}
\beta_p&=\frac{\alpha_{p+j-1}}{\alpha_{p+j}}\beta_{p+j},\quad p\neq k-\frac{j}{2},k,\label{monad_p}\\
\beta_{k-\frac{j}{2}}&=\frac{\alpha_{\frac{j}{2}-1}}{\alpha_{\frac{j}{2}}}\left(\beta_{\frac{j}{2}}-uy\alpha_{\frac{j}{2}-1}^{-1}\right),\label{monad_k-j/2}\\
\beta_k&=\frac{\alpha_{j-1}}{\alpha_j}\left(\beta_j-vz\alpha_{j-1}^{-1}\right).\label{monad_k}
\end{align}
In the case $j=2k$, (\ref{monad_k-j/2}) and (\ref{monad_k}) are replaced by
\begin{align}\label{monad_k_j=2k}
    \beta_k=\frac{\alpha_{k-1}}{\alpha_k}\left(\beta_k-(uy+vz)\alpha_{k-1}^{-1}\right).
\end{align}
Suppose $\beta_r=0$ for some $r\in\{1,\dots,k\}$.
\paragraph{Case 1: $r\neq k-\frac{j}{2},k\mbox{ }(\text{mod }\aleph)$.} From (\ref{monad_p}), this means $\beta_p=0$ for all $p=r\mbox{ }(\text{mod }\aleph)$. Define the vector spaces
$$E_p^\aleph=\mathrm{sp}_\C\{\vec{e}_n\::\:n=p\mbox{ }(\text{mod }\aleph)\},\quad p\in\{1,\dots,\aleph\}.$$
By (\ref{C_k_A_on_basis}) and (\ref{C_k_B_on_basis}), we have $A(E_p^\aleph))\subset E_p^\aleph$ and $B(E_{q-1}^\aleph)=0$, where $q=r\mbox{ }(\text{mod }\aleph)$. As $r\neq k-\frac{j}{2},k$, we also have by (\ref{C_k_D_on_basis}) that $D(E_{q-1}^\aleph)=0$. Therefore, for all $v\in E_{q-1}^\aleph$, and $\Pi\in\II$, we have
$$D\Pi(A,B)v=0,$$
which violates lemma \ref{words_cond} since $E_p^\aleph\neq 0$ for all $p$. Hence, $\beta_r\neq 0$ for $r\neq k-\frac{j}{2},k\mbox{ }(\text{mod }\aleph)$.
\paragraph{Case 2: $r=k-\frac{j}{2}$ $(\text{mod }\aleph)$ or $r=k$ $(\text{mod }\aleph)$.} In this case, the monad equations tell us nothing useful about the other parameters. For this, we rely on the symmetry equations (\ref{C_2k_rot_A})-(\ref{C_2k_rot_D_2}), for which we need a gauge transformation $g\in GL(k,\C)$. A convenient way to write such a matrix is in terms of the `basis' $\{S^p\}_{p=1}^k$:
\begin{align}\label{g_ansatz_detB}
 g=\sum_{a=1}^k\mathrm{diag}\{\gamma_a^1,\dots,\gamma_a^k\}S^a,\quad \gamma_p^q\in\C.
\end{align}
The condition $g\in GL(k,\C)$ implies that for each fixed $p$ and $q$, the $p$-th row and $q$-th column cannot be zero. In the context of (\ref{g_ansatz_detB}), this means we cannot have $\gamma_a^p=0$ for all $a$, or $\gamma_a^{a+q}=0$ for all $a$. Using this $g$ as in (\ref{g_ansatz_detB}), the symmetry equation (\ref{C_2k_rot_A}) implies
\begin{align}\label{gammas_A_rot}
    \gamma_a^p=e^{\imath\frac{j\pi}{k}}\frac{\alpha_{p-a+j}}{\alpha_{p+j}}\gamma_a^{p+j},\quad\text{for all }p=1,\dots,k,
\end{align}
and (\ref{C_2k_rot_C_1})-(\ref{C_2k_rot_D_2}) imply that the only non-zero entry in columns $j$ and $j/2$, and rows $k-j/2-1$ and $k-1$ are given by $\gamma_{k-\frac{j}{2}}^x$ (for the corresponding value of $x$). In other words, for $a\neq k-\frac{j}{2}$
\begin{align}\label{gammas_zero}
\gamma_a^p=0,\quad\text{for all } p\in\left\{a+j,a+\frac{3j}{2},k-\frac{j}{2}-1,k-1\right\}.
\end{align}
Combining (\ref{gammas_A_rot}) and (\ref{gammas_zero}), we have that the same statement as above holds for row $p$ when $p=k-j/2-1,k-1$ $(\text{mod }\aleph)$, and column $q$ when $q=j,j/2\mbox{ }(\text{mod }\aleph)$. In other words, we have
\begin{align}\label{gammas_non_zero_k-j/2}
  \gamma_{k-\frac{j}{2}}^p\neq0,\quad\text{for all } p=k-\frac{j}{2},k,k-\frac{j}{2}-1,k-1\mbox{ }(\text{mod }\aleph).
\end{align}
Using all of this, we may focus attention to the coefficients of $S^{k-\frac{j}{2}}$ in (\ref{C_2k_rot_B}), which translates in this context to the following equations:
\begin{align}
\gamma_{k-\frac{j}{2}}^p\beta_{p+\frac{j}{2}}&=e^{\imath\frac{\pi}{k}}\gamma_{k-\frac{j}{2}}^{p-1}\beta_p,\quad p=1,\dots,k-1,\label{k-j/2_eqn_1tok-1}\\
\gamma_{k-\frac{j}{2}}^k(\beta_{\frac{j}{2}}-uy\alpha_{\frac{j}{2}-1}^{-1})&=e^{\imath\frac{\pi}{k}}\gamma_{k-\frac{j}{2}}^{k-1}\beta_{k}.\label{k-j/2_eqn_k}
\end{align}
By (\ref{gammas_non_zero_k-j/2}), the relevant coefficients $\gamma_{k-\frac{j}{2}}^x$ in (\ref{k-j/2_eqn_1tok-1}) are non-zero for $p=k-\frac{j}{2}$ $(\text{mod }\aleph)$ or $p=k$ $(\text{mod }\aleph)$. So, if $\beta_r=0$ for some $r=k-\frac{j}{2}$ $(\text{mod }\aleph)$ or $r=k$ $(\text{mod }\aleph)$, this implies that $\beta_p=0$ for all $p=r\mbox{ }(\text{mod }\mathrm{gcd}(k,j/2))=0\mbox{ }(\text{mod }\mathrm{gcd}(k,j/2))$. In particular, this means that $\beta_{\frac{j}{2}}=\beta_k=0$, which hence implies by (\ref{k-j/2_eqn_k}) that either $u=0$ or $y=0$, which is a contradiction of lemma \ref{lemma_C,D_evec}. So $\beta_r\neq0$ for all $r=k-\frac{j}{2}\mbox{ }(\text{mod }\aleph)$, or $r=k\mbox{ }(\text{mod }\aleph)$.\\

\noindent Since each case has arrived at a contradiction, we must have that $\beta_r\neq 0$ for all $r=1,\dots,k$, i.e. $\det B\neq 0$.\hfill$\square$\\

\noindent Due to lemma \ref{lemma_detBnon0}, we may now correctly consider a gauge transformation of the form
 \begin{align*}
 h=\mathrm{diag}\left\{\prod_{i=2}^{k-1}\beta_{i},\prod_{i=3}^{k-1}\beta_{i},\dots,\beta_{k-1},1,\prod_{i=1}^{k-1}\beta_i\right\}.
 \end{align*}
After re-labelling the other parameters, this transforms (\ref{ABCD_C_k}) such that $B=\mathrm{diag}\{1,\dots,1,\beta\}S$. We may furthermore set $y=1$ via a subsequent gauge transformation of the form $h'=y\mathbb{1}_k$. This removes all the remaining gauge freedom, and hence fully fixes the gauge equivalence class representatives.

The final task is to take this data and solve the $\rho(C_{2k}^{j,\omega})$-symmetry equations and the monad equations, in line with the full-rank-conditions. If we assume $g$ takes the form as in (\ref{g_ansatz_detB}), then (\ref{C_2k_rot_B}) yields equations of the form
\begin{align}\label{B_symmetry_eqn_k-j/2_1tok-1}
\gamma_i^r=\lambda_i^r\gamma_i^{r-1},\quad\text{for all }i,r=1,\dots,k,
\end{align}
where $\lambda_i^r\neq 0$. In particular, (\ref{B_symmetry_eqn_k-j/2_1tok-1}) implies
\begin{align*}
\gamma_i^r=\left(\prod_{p=0}^{r+1}\lambda_i^{r-p}\right)\gamma_i^{k-1},
\end{align*}
which is $0$ for all $i\neq k-\frac{j}{2}$ by (\ref{gammas_zero}). This means our ansatz for $g$ reduces to
\begin{align}
g=e^{\imath\frac{\pi}{2k}}\mathrm{diag}\{\gamma_1,\dots,\gamma_k\}S^{-\frac{j}{2}},
\end{align}
where we have relabelled the components for simplicity. Using this and the form of $A$, we may solve (\ref{g_G_convert}) to give
\begin{align}\label{alphas_general_A}
 \alpha_{p}=\omega_{p-j}\gamma_{p-j}^{-1}\gamma_{p-\frac{j}{2}}^{-1},\quad p=1,\dots,k.
 \end{align}
A quick check shows that this also satisfies (\ref{C_2k_rot_A}). We may also solve (\ref{C_2k_rot_B})-(\ref{C_2k_rot_D_2}) to yield
 \begin{align}\label{C_2k_monad_matrix_data_solution}
 \begin{array}{cc}
    \gamma_p=\left\{\begin{array}{cl}
        e^{\imath\frac{p\pi}{k}}\gamma_k,&1\leq p\leq k-\frac{j}{2}-1,\\
        2e^{\imath\frac{\left(p-\frac{j}{2}\right)\pi}{k}}u^{-1}\gamma_k^{-1}, & k-\frac{j}{2}\leq p\leq k-1,
    \end{array}\right.&v=e^{\imath\left(\frac{j\pi}{4k}-\omega\right)}\gamma_{\frac{j}{2}}^{-1}u,\\
\beta=\left\{\begin{array}{cc}
\frac{1}{2}u\gamma_k^2e^{\imath\frac{j\pi}{2k}},&j\neq 2k,\\
\frac{1}{2}u\gamma_k^2,&j=2k,
\end{array}\right.&z=e^{\imath\left(\omega+\frac{\pi(3j+4)}{4k}\right)}\gamma_{k-\frac{j}{2}-1},
\end{array}
\end{align}
where the case $j=2k$ is extracted for $\gamma_p$ by taking the first case. This leaves us with two non-zero complex parameters $u$ and $\gamma\equiv\gamma_k$. One may check after a relatively short calculation that these data satisfy the monad equations and full-rank-conditions (\ref{monad_eqn}) and (\ref{full-rank_conditions}). For the latter, it is useful to utilize remark \ref{ABCD,std_basis_remark} and lemma \ref{words_cond}. It is also straight-forward to check that no two solutions are gauge equivalent for varying $u$ and $\gamma$.

Along with the observation that $\MM_k^j(\rho)$ parameterises $\CC_k(\rho(C_{2k}^{j,\omega}))$, all of this completes the proof of theorem \ref{theorem_cyclic_2k_rotation}.\hfill$\square$
\section{Nahm data for symmetric calorons}\label{section_Nahm_calculation}
Despite indicating geometric and topological information about calorons, the monad matrix data lacks the accessibility of Nahm data when it comes to reconstructing the associated caloron. Whilst the Nahm transform is in general a hard process, constructing a caloron from monad matrices is harder still -- to do this one must settle for either reproducing the Nahm data via a Nahm complex, or, as outlined in \cite{CharbonneauHurtubise2008Rat.map}, obtain the caloron from the holomorphic vector bundles over $\CP^1\times\CP^1$ which arise as quotients of the maps (\ref{full-rank_conditions}). Both of these processes require solving several ordinary or partial differential equations.

One is usually able to implicitly derive the caloron via the Nahm transform with quite a lot of accuracy, and in the case $k=1$, this has been done explicitly \cite{KraanVanBaal1998,LeeLu1998}. For higher charge cases, the success has so far been only with numerical methods. For calorons, this has been done only for very special cases \cite{MuranakaNakamulaSawadoToda2011numerical2,Nakamula2014aspects}, but many examples have been considered in the case of monopoles \cite{Sutcliffe1997BPS}. Recent work by Braden and Enolski \cite{bradenenolski2017construction} has shed light on how to construct monopoles explicitly from Nahm data, yet it is still unknown whether these results apply to calorons.

The upshot is, it is significantly more desirable to have Nahm data than monad matrix data when it comes to actually constructing calorons. It would be nice to be able to know the Nahm data for our cyclic calorons from theorem \ref{theorem_cyclic_2k_rotation}, so we shall hence dedicate this final section to the story of this Nahm data.
\subsection{Cyclic Nahm data}
In order to construct fixed points under the action of $\rho(C_{2k}^{j,\omega})$ in the moduli space $\mathcal{N}_k$ of $k$-Nahm data, this amounts to finding smooth functions $g_p:I_p\to U(k)$, and Nahm data $T_p^\lambda:I_p\to\mathfrak{u}(k)$, $(u_p,w_p)\in(\C^k\times\C^k)\setminus\{(0,0)\}$, satisfying
\begin{align}
    \begin{array}{rcl}
         T_p^0(s)&=&\left(g_qT_q^0g_q^{-1}-\dfrac{dg_q}{ds}g_q^{-1}\right)\left(s+(-1)^q\frac{\mu_0}{2}\right)+\dfrac{\imath j\pi}{k\mu_0}\mathbb{1},\\
         T_p^1(s)&=&\left(g_qT_q^1g_q^{-1}\right)\left(s+(-1)^q\frac{\mu_0}{2}\right),\\
         T_p^2(s)&=&\left(g_q\left(\cos\left(\frac{\pi}{k}\right)T_q^2+\sin\left(\frac{\pi}{k}\right)T_q^3\right)g_q^{-1}\right)\left(s+(-1)^q\frac{\mu_0}{2}\right),\\
         T_p^3(s)&=&\left(g_q\left(-\sin\left(\frac{\pi}{k}\right)T_q^2+\cos\left(\frac{\pi}{k}\right)T_q^3\right)g_q^{-1}\right)\left(s+(-1)^q\frac{\mu_0}{2}\right),\\
         (u_1,w_1)&=&e^{\imath\omega}g_1(-\mu_0/4)\left(e^{-\imath\frac{\pi}{2k}}u_2,e^{\imath\frac{\pi}{2k}}w_2\right),\\
         (u_2,w_2)&=&e^{-\imath\omega}g_1(\mu_0/4)\left(e^{-\imath\frac{\pi}{2k}}u_1,e^{\imath\frac{\pi}{2k}}w_1\right),
    \end{array}\label{Cyclic_rot_2k_eqns_Nahm}
\end{align}
where $q=p+1\mbox{ }(\text{mod }2)$.

To solve this system, in analogy to the methods for monad matrices outlined and executed in the previous section, we may first consider Nahm matrices on $I_1$ with $C_k^{j,0}$-symmetry, that is, such that
\begin{align}
    \begin{array}{rcl}
    T_1^0(s)&=&G(s)T_1^0G(s)^{-1}-\frac{dG}{ds}(s)G(s)^{-1}+\frac{2\imath j\pi}{k\mu_0}\mathbb{1},\\
    T_1^1(s)&=&G(s)T_1^1(s)G(s)^{-1},\\
    T_1^2(s)&=&G(s)\left(\cos\left(\frac{2\pi}{k}\right)T_1^2(s)+\sin\left(\frac{2\pi}{k}\right)T_1^3(s)\right)G(s)^{-1},\\
    T_1^3(s)&=&G(s)\left(-\sin\left(\frac{2\pi}{k}\right)T_1^2(s)+\cos\left(\frac{2\pi}{k}\right)T_1^3(s)\right)G(s)^{-1},
    \end{array}\label{Cyclic_k_eqns_Nahm}
\end{align}
for some $G:I_1\to U(k)$, and data $T_1^\lambda:I_1\to\mathfrak{u}(k)$ satisfying Nahm's equations (\ref{Nahm_eqn}) on $I_1$. This $G$ is to be related to the $g_p$ found in (\ref{Cyclic_rot_2k_eqns_Nahm}) via
\begin{align}\label{g_convert_G_Nahm}
    G(s)=g_2(s+\mu_0/2)g_1(s).
\end{align}
Once we have such data and gauge transformations, we may fix the form of the matching data $(u_p,w_p)$ as in (\ref{Cyclic_rot_2k_eqns_Nahm}), and use $g_1$ to set $T_2^\alpha:I_2\to\mathfrak{u}(k)$ as
\begin{align}\label{2nd_interval_Nahm_data}
\begin{array}{rcl}
T_2^0(s)&=&\left(g_1T_1^0g_1^{-1}-\dfrac{dg_1}{ds}g_1^{-1}\right)\left(s-\frac{\mu_0}{2}\right)+\dfrac{\imath j\pi}{k\mu_0}\mathbb{1},\\
         T_2^1(s)&=&\left(g_1T_1^1g_1^{-1}\right)\left(s-\frac{\mu_0}{2}\right),\\
         T_2^2(s)&=&\left(g_1\left(\cos\left(\frac{\pi}{k}\right)T_1^2+\sin\left(\frac{\pi}{k}\right)T_1^3\right)g_1^{-1}\right)\left(s-\frac{\mu_0}{2}\right),\\
         T_2^3(s)&=&\left(g_1\left(-\sin\left(\frac{\pi}{k}\right)T_1^2+\cos\left(\frac{\pi}{k}\right)T_1^3\right)g_1^{-1}\right)\left(s-\frac{\mu_0}{2}\right).
\end{array}
\end{align}
It is straight-forward to check that (\ref{2nd_interval_Nahm_data}) along with the symmetric matching data, and the $T_1^\lambda$ satisfying (\ref{Cyclic_k_eqns_Nahm}) yields a general solution to (\ref{Cyclic_rot_2k_eqns_Nahm}).

The final obstruction is solving the matching conditions (\ref{matching_cond}) with the symmetric data. This only needs to be performed at $s=\mu_0/4$, as the other condition is equivalent due to the symmetric form.
\subsection{The case \texorpdfstring{$k=2$}{k=2}}
We shall present a prototype example of $\rho(C_{2k}^{j,\omega})$-symmetric Nahm data in the case $k=2$. In accordance with lemma \ref{lemma_j_even}, the only actions in the circle translations to consider are $j=4\equiv0$ and $j=2$.

It turns out that all solutions in the case $k=2$, $j=0$, are known. These are described by the `crossed solution' of N\'ogr\'adi and coworkers. We will describe this data in the next section. In the remaining sections we shall construct the previously unknown solutions in the case $k=j=2$, which we call the \textit{oscillating solution}.
\subsubsection{The crossed solutions}
In an article by Bruckmann et. al. \cite{BruckmannNogvanB2003constituent}, a general, unmatched solution to Nahm's equations (\ref{Nahm_eqn}) was given in the case of charge $(2,2)$. This fixes the Nahm matrices $T_p^\lambda$ in terms of three functions satisfying the differential equations $f_1'=-f_2f_3$ and cyclic, which come from the Nahm equations (in \cite{BruckmannNogvanB2003constituent}, the equations considered are the \textit{self-dual} Nahm equations, so the minus sign is omitted). These are solved by the elliptic functions
\begin{align}\label{diff_eqn_soln} \Phi_1(s)=-\frac{\kappa'}{\cn_{\kappa}(s)},\quad \Phi_2(s)=\frac{\kappa'\sn_\kappa(s)}{\cn_\kappa(s)},\quad \Phi_3(s)=\frac{\dn_{\kappa}(s)}{\cn_{\kappa}(s)}.
\end{align}
Here the parameter $\kappa\in[0,1]$ is the \textit{elliptic modulus}, and $\kappa'=\sqrt{1-\kappa^2}$. For details see \cite{AbramowitzStegun1964}. Our notation is slightly different, but may be translated simply via $\kappa\leftrightarrow\sqrt{m}$. We will also often adopt Glaisher's notation for quotients, for example $\sn(s)/\cn(s)\equiv\scn(s)$, etc.

Note that the minus sign in (\ref{diff_eqn_soln}) could have been chosen to be attached to any of the three functions in a gauge equivalent way. Three families of solutions are described in \cite{BruckmannNogvanB2003constituent}, of which two are shown to have axial symmetry in \cite{Harland2007}. The remaining solutions are the `crossed' solutions, described most clearly by N\'ogr\'adi in \cite{NogradiThesis}. One may show that these families have $\rho(C_4^{0,\omega})$-symmetry around the axis $x^2$.

With our convention of fixing the $x^1$ axis, we may describe the crossed solutions as follows. The Nahm matrices on $I_1$ are given by
\begin{align}\label{Nahm_crossed_I_1}
\begin{array}{c}
\begin{array}{cc}
    T_1^0=\dfrac{\imath}{\mu_0}\theta\mathbb{1},&
    T_1^1(s)=\imath\left(\alpha\mathbb{1}+\dfrac{D}{2}\kappa'\scn_\kappa(Ds)\sigma^3\right),\end{array}\\
    T_1^2(s)=\imath\dfrac{D}{2}\left(\cos\phi\kappa'\nc_\kappa(Ds)\sigma^1+\sin\phi\dc_\kappa(Ds)\sigma^2\right)\\
    T_1^3(s)=\imath\dfrac{D}{2}\left(\cos\phi\dc_\kappa(Ds)\sigma^2-\sin\phi\kappa'\nc_\kappa(Ds)\sigma^1\right),
\end{array}
\end{align}
and the matrices on $I_2$ are determined from (\ref{Nahm_crossed_I_1}) as in (\ref{2nd_interval_Nahm_data}) with the gauge transformation
\begin{align*}
    g_p(s)=\frac{\sqrt{2}}{2}(\imath\mathbb{1}-\sigma^3).
\end{align*}
The matching data for the crossed solutions are given by the vectors
\begin{align}\label{matching_data_crossed}
(u_1,w_1)=\left(\begin{pmatrix}
0\\
\lambda
\end{pmatrix},\begin{pmatrix}
\lambda e^{\imath\left(\phi+\frac{\pi}{2}\right)}\\0
\end{pmatrix}\right),\quad
(u_2,w_2)=\left(\begin{pmatrix}
0\\
\lambda e^{-\imath\omega}
\end{pmatrix},\begin{pmatrix}
\lambda e^{\imath\left(\phi+\frac{\pi}{2}-\omega\right)}\\
0
\end{pmatrix}\right).
\end{align}
One can straight forwardly check that all of this solves (\ref{Cyclic_rot_2k_eqns_Nahm}) in the case $n=k=2$, $j=0$, and are hence $\rho(C_4^{0,\omega})$-symmetric, moreover, they give all $(2,2)$-calorons with such symmetry. The data (\ref{Nahm_crossed_I_1}) and (\ref{matching_data_crossed}) are required to obey the matching conditions (\ref{matching_cond}), which here reduce to
\begin{align*}
    2D\kappa'\scn_\kappa(D\mu_0/4)&=\lambda^2,\\
    D(\dc_\kappa(D\mu_0/4)-\kappa'\nc_\kappa(D\mu_0/4))&=\lambda^2.
\end{align*}
The free parameters in this solution are given by $\phi\in[0,2\pi)$, $\theta\in[0,4\pi)$, $\alpha\in\R$, and $\kappa\in(0,1)$. One may show that for all $\kappa\in(0,1)$, there exists a unique solution $(D_\kappa,\lambda_\kappa)$ to the matching conditions in terms of $\kappa$ (the analysis is similar to the detailed example we present in the next section).

It is worth remarking that this solution family is called `crossed' as in a particular limit where the caloron becomes abelian \cite{NogradiThesis}, the constituent monopoles localise as two crossed ellipses occupying the same plane.
\subsubsection{The oscillating solutions}
In this and the remaining sections, we construct the Nahm data for $\mathcal{C}_2(\rho(C_4^{2,\omega}))$. Our ansatz is given by solving (\ref{Cyclic_k_eqns_Nahm}), namely
\begin{align}
\arraycolsep=1pt\def\arraystretch{2.2}
\begin{array}{l}
T_1^0=\dfrac{\imath}{\mu_0}\theta\mathbb{1},\quad
T_1^1(s)=\imath\left(\alpha\mathbb{1}+\dfrac{D}{2}f_1(Ds)\sigma^3\right),\\
T_1^2(s)=\imath\dfrac{D}{2}\left(\cos\phi f_2(Ds)\sigma^2+\sin\phi f_3(Ds)\sigma^1\right),\\
T_1^3(s)=\imath\dfrac{D}{2}\left(\cos\phi f_3(Ds)\sigma^1-\sin\phi f_2(Ds)\sigma^2\right),
\end{array}\label{C_2^2,omega_ansatz}
\end{align}
with gauge transformation $G(s)=e^{\imath\frac{2\pi}{\mu_0}s}\sigma^3$. The parameters $\theta,\alpha,\phi,$ and $D$ are constants, and the functions $f_p$ are subject to a differential equation obtained from Nahm's equations, which in this case reduce to the \textit{Euler top equations} $f_j'=-f_kf_l$ (and cyclic). This has general solution given by the functions $\Phi_1,\Phi_2,\Phi_3$ as in (\ref{diff_eqn_soln}).

The next step is to solve the equation (\ref{g_convert_G_Nahm}). In general, this factorisation problem is potentially quite hard, however, we can use the intended symmetry to our advantage. Analysing the holonomy of $T^0$ under the action of a circle translation, and fixing matching data with $\rho(C_4^{2,\omega})$-symmetry, leads us to fixing a gauge such that
\begin{align}\label{g_p_unsymm}
g_1(s)=g_2(s)=\begin{pmatrix}
0&1\\
e^{-\imath\psi}e^{\imath\frac{2\pi}{\mu_0}s}&0
\end{pmatrix}e^{\imath\psi},
\end{align}
for some $\psi\in\R$. These may easily be checked to satisfy $G(s)=g_2(s+\mu_0/2)g_1(s)$. The symmetric matching data are given by
\begin{align}\label{C_4^2,omega_matching_data}
(u_1,w_1)=\left(\begin{pmatrix}
u\\
0
\end{pmatrix},\begin{pmatrix}
0\\
w
\end{pmatrix}\right),\quad (u_2,w_2)=\left(\begin{pmatrix}
0\\
e^{\imath\left(\frac{\pi}{4}-\psi-\omega\right)}u
\end{pmatrix},\begin{pmatrix}
e^{\imath\left(\frac{\pi}{4}+\psi+\omega\right)}w\\0
\end{pmatrix}\right).
\end{align}
Setting the Nahm matrices on $I_2$ according to (\ref{2nd_interval_Nahm_data}), we can obtain a complete family of unmatched $\rho(C_4^{2,\omega})$-symmetric Nahm data, satisfying (\ref{Cyclic_rot_2k_eqns_Nahm}) (for $n=k=j=2$) with gauge transformations (\ref{g_p_unsymm}).

\subsubsection{Solving the matching conditions}
The matching conditions for the data above may be written as
\begin{align}
-D\left(f_1\left(D\frac{\mu_0}{4}\right)+f_1\left(-D\frac{\mu_0}{4}\right)\right)=
    |u|^2&=|w|^2,\label{matching_circle_symm_1}\\
e^{-\imath\phi}\dfrac{D}{2}\left(f_2\left(D\frac{\mu_0}{4}\right)-f_3\left(D\frac{\mu_0}{4}\right)+e^{2\imath\psi}\left(f_2\left(-D\frac{\mu_0}{4}\right)+f_3\left(-D\frac{\mu_0}{4}\right)\right)\right)&=u\ol{w},\label{matching_circle_symm_2}\\
D\left(f_3\left(D\frac{\mu_0}{4}\right)+f_2\left(D\frac{\mu_0}{4}\right)-e^{-2\imath\psi}\left(f_2\left(-D\frac{\mu_0}{4}\right)-f_3\left(-D\frac{\mu_0}{4}\right)\right)\right)&=0,\label{matching_circle_symm_3}.
\end{align}
Straight-forward analysis of these equations reveals that we may set
\begin{align}
u=\lambda,\quad w=\lambda e^{\imath(\phi+n\pi)},
\end{align}
for $\lambda\in\R^+$ and $n\in\Z$. Furthermore, these reveal that $f_2$ and $f_3$ must have opposite parity, and $f_1$ must be even. We may, without loss of generality, fix $f_2$ even, and $f_3$ odd, as the opposite case is gauge equivalent. It then follows from (\ref{matching_circle_symm_2}) and (\ref{matching_circle_symm_3}) that $2\psi=2r\pi$, with different values of $r$ giving gauge equivalent data. This means that the matching equations (\ref{matching_circle_symm_2}) and (\ref{matching_circle_symm_3}) are reduced to the cases of finding solutions to the following equations:
\begin{align}\label{Aa}
(-1)^n\left(\kappa'\scn_\kappa(x)-\dc_\kappa(x)\right)+2\kappa'\nc(x)=0,
\end{align}
or
\begin{align}\label{Ab}
(-1)^n\kappa'\left(\scn_\kappa(x)+\nc_\kappa(x)\right)-2\dc_\kappa(x)=0,
\end{align}
where we have relabelled $x\equiv D\mu_0/4$. We must check for which values of $x,\kappa,n$ these equations have solutions, if any exist at all. In appeal to theorem \ref{theorem_cyclic_2k_rotation}, we expect there to be a unique family of solutions. We shall proceed to analyse existence or non-existence of solutions to (\ref{Aa}) and (\ref{Ab}).

Note that each of the $\Phi_j$ have poles at $x=rK$, for odd integers $r$, where $K(\kappa)\in\R$ is the \textit{complete elliptic integral of the first kind}
\begin{align*}
    K(\kappa)=\int_0^{\frac{\pi}{2}}\dfrac{1}{\sqrt{1-\kappa^2\sin\theta}}d\theta.
\end{align*}
It is possible, due to the residues of the individual functions, that $rK$ could be a solution to (\ref{Aa}) or (\ref{Ab}). However, an important property of Nahm data is that it needs to be bounded and continuous, and hence pole-free across all of $I_p$. Therefore, we can throw out any solutions which lead to the arguments $Ds$ attaining poles for some $s\in I_1$. A straight-forward calculation reveals that the only values of $x=D\mu_0/4$ which lead to pole-free data are found in $-K<x<K$, hence we shall fix our attention to this interval only.
\begin{lemma}\label{eqAb_non}
There are no solutions to equation (\ref{Ab}) in the range $-K<x<K$, for all $\kappa,n$.
\end{lemma}
\textit{Proof}. Multiplying through by $\cn_\kappa(x)$, and rearranging using elliptic function identities, (\ref{Ab}) implies
\begin{align*}
    \left(4-3\kappa'^2\right)\sn_\kappa^2(x)+2\kappa'^2\sn_\kappa(x)+\kappa'^2-4&=0.
\end{align*}
Since $\kappa'\in[0,1]$, we have $4-3\kappa'^2>0$, so that this is solved if and only if
\begin{align}\label{eqnAb_sn_sol}
\sn_\kappa(x)=1,\quad\text{or}\quad\sn_\kappa(x)=\dfrac{\kappa'^2-4}{4-3\kappa'^2}.
\end{align}
Note also
$$\dfrac{\kappa'^2-4}{4-3\kappa'^2}\leq\dfrac{3\kappa'^2-4}{4-3\kappa'^2}=-1.$$
As $\sn_\kappa(x)\in[-1,1]$, this means (\ref{eqnAb_sn_sol}) is satisfied only when $\sn_\kappa(x)=\pm1$, that is, when $x=K(\kappa)$ (modulo $4K$), which is not in the interval $(-K,K)$.\hfill$\square$\\

\noindent As stipulated above, we may only look for solutions in the interval $(-K,K)$. Thus, by Lemma \ref{eqAb_non}, the only remaining case to consider is (\ref{Aa}), which corresponds to
$$f_1(s)=-\kappa'\nc_\kappa(s),\quad f_2(s)=\dc_\kappa(s),\quad f_3(s)=\kappa'\scn_\kappa(s).$$
\subsubsection*{Solutions to (\ref{Aa})}
In this section, we consider the equation (\ref{Aa}). First we shall show that no solutions exist when $n=1\mbox{ }(\text{mod }2)$, and then show existence of a unique family of solutions in the case $n=0\mbox{ }(\text{mod }2)$ within the required range. Consider the functions
\begin{align}
F_1(x,\kappa)&=\kappa'\scn_\kappa(x)-\dc_\kappa(x),\\
F_2(x,\kappa)&=2\kappa'\nc_\kappa(x).
\end{align}
There are no solutions to the equation $F_1(x,\kappa)=F_2(x,\kappa)$ since in the range $-K<x<K$, we have
$$F_1\leq-1<2\kappa'\leq F_2.$$
This rules out the case $n=1$, and hence we are only left with the case $n=0$, for which we are interested in finding solutions to $F_1=-F_2$. We shall soon show that there exists a subset $U\subset[0,1]$ such that there exist solutions $x\in(-K,K)$ to $F_1(x,\kappa)=F_2(x,\kappa)$ for all $\kappa\in U$. However, in order for these solutions to yield Nahm data, they must solve not only this equation, but the complete set of matching conditions. Other than the equation we're considering, the other dependent is (\ref{matching_circle_symm_1}), which in our situation gives the condition
\begin{align}\label{positive_nc}x\nc_\kappa(x)>0.\end{align}
As $\nc_\kappa(x)>0$ for all $-K<x<K$, we easily see that (\ref{positive_nc}) is satisfied only if $0<x<K$. This all leads to the following.

\begin{proposition}\label{prop_existence_soln_Nahm}
For $0<x<K$, there exists a unique solution to the equation $F_1(x,\kappa)=-F_2(x,\kappa)$ if and only if $\kappa\in\left(\frac{\sqrt{3}}{2},1\right)$.
\end{proposition}
\textit{Proof}. By rearranging and squaring, and using elliptic function identities, it is straight-forward to see that any solution to $F_1=-F_2$ necessarily satisfies
$$\sn_\kappa^2(x)+4\kappa'^2\sn_\kappa(x)+4\kappa'^2-1=0,$$
which is solved if and only if
\begin{align}\label{solution_matching_nec}
    \sn_\kappa(x)=-1,\quad\text{or}\quad\sn_\kappa(x)=1-4\kappa'^2=4\kappa^2-3.
\end{align}
As $\sn_\kappa(x)\in(0,1)$ for all $x\in(0,K)$, and $\kappa\in[0,1]$, (\ref{solution_matching_nec}) is solved in this range only if $\kappa\in\left(\frac{\sqrt{3}}{2},1\right)$.

Conversely, let $\kappa\in\left(\frac{\sqrt{3}}{2},1\right)$, and consider the function $F(x,\kappa)=F_1(x,\kappa)+F_2(x,\kappa)$. As $\kappa\in\left(\frac{\sqrt{3}}{2},1\right)$, we have $\kappa'\in\left(0,\frac{1}{2}\right)$, thus
\begin{align}\label{F(0)}
F(0,\kappa)=2\kappa'-1<0.
\end{align}
Also, $F$ has a simple pole with residue $2$ at $x=K$, so $F(K,\kappa)=\infty>0$. As $F$ is continuous on $(0,K)$, combining this with (\ref{F(0)}), the intermediate value theorem says that there exists $x_0\in(0,K)$ such that $F(x_0,\kappa)=0$, that is $F_1(x_0,\kappa)=-F_2(x_0,\kappa)$. Moreover, this solution $x_0$ will be unique as both $F_1'>0$ and $F_2'>0$ for all $0<x<K$, and hence the same for $F$.\hfill$\square$
\subsubsection{The invariant Nahm data}
Following the analysis of the matching conditions in the previous section, we conclude that there is only one family of $\rho(C_4^{2,\omega})$-symmetric Nahm data. This is given by the Nahm matrices on $I_1$ (\ref{C_2^2,omega_ansatz}) with the replacements
\begin{align}
    f_1=-\kappa'\nc_\kappa,\quad f_2=\dc_\kappa,\quad f_3=\kappa'\scn_\kappa,\quad D=D_\kappa,
\end{align}
which determine the Nahm matrices on $I_2$ via (\ref{2nd_interval_Nahm_data}) using the gauge transformation (\ref{g_p_unsymm}) with $\psi=0$. The matching data are
\begin{align}
(u_1,w_1)&=\left(\begin{pmatrix}
\sqrt{2D_\kappa\kappa'\nc\left(D_\kappa\frac{\mu_0}{4}\right)}\\
0
\end{pmatrix},\begin{pmatrix}
0\\
\sqrt{2D_\kappa\kappa'\nc\left(D_\kappa\frac{\mu_0}{4}\right)} e^{\imath(\phi+\pi)}
\end{pmatrix}\right),\label{C_4^2,omega_matching_data_soln_1}\\ (u_2,w_2)&=\left(\begin{pmatrix}
0\\
e^{\imath\left(\frac{\pi}{4}-\omega\right)}\sqrt{2D_\kappa\kappa'\nc\left(D_\kappa\frac{\mu_0}{4}\right)}
\end{pmatrix},\begin{pmatrix}
e^{\imath\left(\frac{3\pi}{4}+\phi+\omega\right)}\sqrt{2D_\kappa\kappa'\nc\left(D_\kappa\frac{\mu_0}{4}\right)}\\0
\end{pmatrix}\right),\label{C_4^2,omega_matching_data_soln_2}
\end{align}
By performing a gauge transformation of the form $h=\mathrm{diag}\{1,\imath\}$, this data satisfies the reality condition $T^\lambda(-s)=T^\lambda(s)^t$, a condition required by the equivalent $Sp(1)$ formalism. The free parameters are $\theta,\phi,\alpha,\kappa$. Up to gauge equivalence, $\theta\in[0,4\pi)$, $\phi\in[0,2\pi)$, $\alpha\in\R$, and $\kappa\in(\sqrt{3}/2,1)$ determines $D_\kappa\in(0,K(\kappa))$ in line with proposition \ref{prop_existence_soln_Nahm}. We call this solution `oscillating' as it is much like the crossed solution, but the crossing of the monopole constituents is shifted temporally by half the period, as opposed to all time.
\section{Summary and concluding remarks}
The purpose of this article was to determine whether families of calorons exist with rotation map symmetry. This endeavour has been successful: we have fully classified the fixed point sets of cyclic groups of order $2k$, whose generator involves the rotation map, in the moduli space of $(k,k)$ calorons with equal monopole masses. From the point of view of explicit construction, our work has made use of the monad matrix data, an approach which to date has not been utilised before in the context of calorons. Additionally, in the case $k=2$, we have derived a new solution to Nahm's equations for one family of these symmetric calorons, and in doing so, this has completed the list of rank $2$ Nahm data with these symmetries.

Regarding the main classification result (theorem \ref{theorem_cyclic_2k_rotation}), an important observation is that the geometry of the fixed point sets is not dependant on $k$, indeed they are all equal to $(\C^\ast)^2$. Interestingly, this is a specific caloron moduli space, namely the space of \textit{unframed, centred,} $(1,1)$-calorons, that is, $\CC_1$ modulo spatial translations and phases. The harder problem of classifying all $(k,k)$-calorons with rotation cyclic symmetry of order $2n$, where $n$ divides $k$ still remains unfinished. What would be interesting to see is whether a similar pattern emerges as in the case $n=k$, for example, do the fixed point sets depend on $k$, and are they all special types of caloron moduli spaces? It is straight-forward to see that no rotation map symmetric calorons exist for $n>k$.

Another question that is left open is how these calorons relate to other solitons. In line with the work in \cite{HarlandWard2008chains}, one expects some of these calorons to provide approximations to symmetric skyrmion chains by taking holonomies along a line in $\R^3$. Another consideration is to instead take the holonomy around the $S^1$, and this is a topic we are currently working on. It would also be interesting to better understand the large scale limits, as in \cite{Harland2007}, as a way of relating to monopoles and calorons of other charges.

Finally, a broader understanding of which rotation map symmetry groups yield symmetric calorons is still largely unknown. Unfortunately, the methods involving the monad matrices, presented in this paper, can only help in the case where the euclidean actions fix the $x^1$ axis, and not generically. The only remaining symmetries that may be investigated in this manner are dihedral reflections, and reflections of $S^1\times\R^3$ of the form $(t,\vec{x})\mapsto(-t,-\vec{x})$, but this we are yet to look into. In terms of the constituent monopole viewpoint, $(k,k)$-calorons bear resemblance to a pair of charge $k$ monopoles. Heuristically, one might expect if there exists a charge $k$ monopole, symmetric with respect to some order $n$ subgroup $H$ of $O(3)$ (with accompanying phase), then by coupling this with the rotation map, there could be calorons symmetric with respect to a corresponding group of order $2n$, which incorporates $H$ and $\rho$. Whilst this has proven to be true in the case of cyclic calorons, it is not necessarily as simple as stipulated here in general. In order to guarantee that the monopoles are constituents for a caloron, they need to \textit{match} correctly, that is, the Nahm data need to solve the matching conditions (\ref{matching_cond}). An analogous conjecture may be formulated in the case of $SU(N)$ calorons for groups of order $Nn$. All of this has not been investigated in any serious detail, but it would certainly be interesting to see how many other symmetric monopoles, if any, correspond to rotation map symmetric calorons.
\section*{Acknowledgements}
I would like to thank Derek Harland for giving me the inspiration to study this topic, and for numerous useful discussions, especially regarding the definitions and the derivations of results.

\medskip

\printbibliography

\end{document}